\documentclass[aps,pre,reprint,groupedaddress,showpacs]{revtex4-1}
\usepackage{amssymb}
\usepackage{mathrsfs}
\usepackage{amsmath}
\usepackage{latexsym}
\usepackage{amsfonts}
\usepackage{graphicx}
\usepackage{float}
\usepackage{amsmath}
\usepackage{mathcomp}
\usepackage{epsfig}
\usepackage{color}
\usepackage{hyperref}
\usepackage{epstopdf}
\begin{document}
\title{Role of interactions and correlations on collective dynamics of molecular motors along parallel filaments}
\author{Tripti Midha}
\author{Arvind
kumar Gupta}
\email[]{akgupta@iitrpr.ac.in}
\affiliation{Department of Mathematics, Indian Institute of Technology Ropar, Rupnagar-140001, Punjab, India.}

\begin{abstract}
Cytoskeletal motors known as motor proteins are molecules that drive cellular transport along several parallel cytoskeletal filaments and support many biological processes.  Experimental evidence suggests that they interact with the nearest molecules of their filament while performing any mechanical work. To understand such mechanism theoretically, a new version of two-channel totally asymmetric simple exclusion process which incorporates interactions in a thermodynamically consistent way is introduced. As the existing approaches for multi-channel systems deviate from analyzing the combined effect of inter and intra-channel interactions, a new approach known as modified vertical cluster mean field is developed. The approach along with monte-carlo simulations successfully encounters some correlations and computes the complex dynamic properties of the system.
Role of symmetry of interactions and inter-channel coupling is observed on the triple points and the particle maximal current. Surprisingly, for all values of coupling rate and most of the interaction splittings, the optimal interaction strength corresponding to maximal current belongs to the case of weak repulsive interactions. Moreover, for weak interaction splittings and with an increase in the coupling rate, the optimal interaction strength tends towards the known experimental results. Coupling in between the lanes decreases the correlations. They are found to be short-range and weaker for repulsive and weak attractive interactions, while long-range and stronger for large attractions.
\end{abstract}

\pacs{05.60.-k, 02.50.Ey, 64.60.-i, 05.70.Ln}
\keywords{ stochastic particle dynamics (theory), molecular motors (theory), correlations (theory), interactions (theory)}
\maketitle

\section{Introduction}
In all living organisms starting from a unicellular yeast to multicellular humans, cells are the most elementary and complex structures which undergo many vital functions such as cell division, gene replication,  cellular transport, cell motility and signaling \cite{alberts2002molecular,bray2001cell,howard2001mechanics}.
 Generally, these processes are actively carried out by enzymatic molecules called motor proteins or molecular motors \cite{kolomeisky2007molecular,chowdhury2013stochastic,kolomeisky2013motor}. In cellular transport, they utilize chemical energy, released from chemical reactions that they catalyze such as hydrolysis of ATP, to deliver cargoes by their active movement along cytoskeletal filaments \cite{kolomeisky2015motor}.
 The proper functioning of motor protein transport is crucial for the cell's survival. Motor proteins' mutation and disruption in cellular transport can contribute to the development of diseases like Alzheimer, hearing loss, virus transport, neurodegenerative and polyestric kidney diseases \cite{schliwa2003molecular}. Recently, several \textit{in vivo} and \textit{in vitro} single-molecular motor experiments have provided a good insight over mechanochemical properties of motor proteins \cite{chowdhury2013stochastic,kolomeisky2013motor,veigel2011moving}.
 However, motor proteins work in a larger team \cite{ally2009opposite},
 and their collective behavior is not  yet well understood \cite{driver2011productive,driver2010coupling,neri2013exclusion}.

Many experiments on kinesin motor proteins reveal that in the presence of neighboring motors, they remain attached to the microtubule for a longer time which results in the formation of clusters \cite{roos2008dynamic}. These interactions are estimated within an energy range $(1.6\pm 0.5)$ $k_BT$ \cite{roos2008dynamic,seitz2006processive}, where $k_B$ is the Boltzmann's constant and $T$ is the thermodynamic temperature. One can assume similar kind of interactions among other cytoskeletal motors such as myosin and dynein, which are also  involved in cellular transport. It is expected that these interactions alter various chemical transitions such as binding and unbinding, backward stepping, forward moving, hydrolysis, etc occurring at a microscopic level. These changes affect the mechano-chemistry and hence collective dynamics of motors. However, the influence of these interactions on the collective transport of motor proteins needs much more investigation. It is thus significantly important to combine the microscopic properties of motor proteins with their collective transport motion  \cite{kolomeisky2015motor,uppulury2012interplay}.

Totally Asymmetric Simple Exclusion Process (TASEP) has become a paradigmatic model to study the collective motion of active particles that hop stochastically and uni-directionally along a linear segment obeying the hard-core exclusion principle. TASEP was first introduced to model mRNA translation by ribosomes \cite{macdonald1968kinetics}. Since then, the process has been successfully applied to understand the collective properties of many particle systems such as traffic flow, protein synthesis, intracellular processes, gel electrophoresis, etc \cite{belitsky2001cellular,chowdhury2008traffic,widom1991repton}. TASEP has also provided insights into the dynamic properties of interacting \cite{campas2006collective,klumpp2004phase,pinkoviezky2013modelling,slanina2008interaction,teimouri2015theoretical,celis2015correlations,hao2016exponential,hao2016theoretical} as well as non-interacting motor proteins \cite{neri2013exclusion,chou2011non,dong2012entrainment}.

 Recently, a new class of single-channel open TASEP has been introduced, which provides a quantitative description for chemical transitions among motor proteins using fundamental thermodynamic concepts \cite{teimouri2015theoretical,celis2015correlations}. The simple mean field theory which completely ignores particle-particle correlations fails to capture the effect of interactions on one lane TASEP model.
It suggests that interactions bring correlations into the system.
 To deal with such correlations several mean field approaches such as cluster mean field theory \cite{teimouri2015theoretical,hao2016theoretical} and modified cluster mean field theory \cite{celis2015correlations} have been proposed. These theories successfully capture the effect of interactions but applicable only to a single filament.

Examining the realistic features inside the cell, where several channels are offered to cellular transport, particles of different channels interact when they are hindered in their pathway \cite{neri2013exclusion}.
It is expected that in presence of multi-channels, the interactions present in between the molecular motors, affects the collective behavior of motor proteins \cite{shaebani2014anomalous}. 
 In the past, original two-channel TASEP system (without interactions) and its variants have been thoroughly explored under different coupling environment with approximation theories such as vertical cluster mean field and continuum limit of mean field equations \cite{pronina2004two,pronina2006asymmetric,juhasz2007weakly,popkov2004hydrodynamic,gupta2013coupling}. However, these studies can not capture correlations because of intra-channel interactions. The available approximation theories which capture such correlations are limited to one channel \cite{teimouri2015theoretical,celis2015correlations,hao2016theoretical} and can not be simply extended to a multi-channel system. 

 In this paper, an effort has been made to analyze the collective dynamics of interacting molecular motors moving on a symmetrically coupled parallel filaments. The inter-channel interactions are incorporated in the original two-channel open TASEP model by modifying its simple transitions rules in a fundamental thermodynamic procedure. We develop a theory called modified vertical cluster mean field (MVCMF) that considers some correlations and calculates the complex properties of an interactive two-channel coupled system.
We find that coupling in between the lanes decreases the correlations; however, they are stronger for large attractive interactions and weaker for repulsive interactions. We observe the combined effect of symmetry of interactions and coupling on particles' collective dynamics. The method can be generalized to more than two lanes and to consider the role of stochastic open attachment and detachment of particles.
\section{Theoretical Description}
\subsection{Model}
The model defines multi-particle motion on a two-channel (lane) lattice each with $N$ $( N\gg1)$ sites to mimic the transport of molecular motors along parallel cytoskeletal filaments. Each lattice site can be occupied by at most one particle under the hard-core exclusion principle. The state of occupancy of the $i^{th}$ site $(1 \leq i \leq N)$ of the lane $l$  (1 or 2) is characterized by an occupation variable $\tau_{i,l}$, where $\tau_{i,l}$ = 0 denotes its empty state while $\tau_{i,l}$ = 1 represents its occupied state.

Besides the exclusion principle, particles in a channel can interact horizontally with its nearest neighbors via energy $E$ associated to the bond connecting two neighboring particles. It can be said that the horizontal hopping in both channels takes place under a short-range interactive strength $E$ or is driven by an external field $E$ \cite{teimouri2015theoretical,celis2015correlations,hao2016exponential}. 

\begin{figure*}
\includegraphics[trim = 40 625 50 40,clip,width=20cm,height=4.5cm]{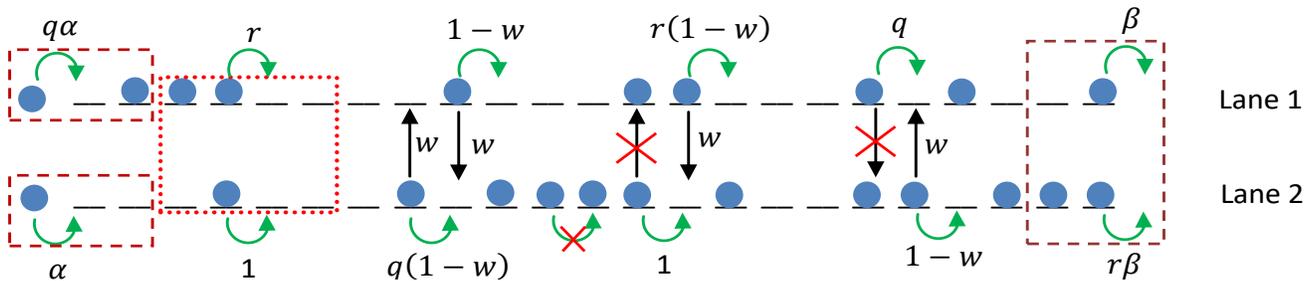}
\caption{Schematic view of the two-channel symmetrically coupled interacting TASEP model. The leftmost and rightmost boxes, respectively, represent the different possibilities for a particle to enter and leave the system. All hopping rates are defined for the system and are irrespective of each lane.}\label{fig1}
\end{figure*}

Moreover, the horizontal transition rules for a particle at $i^{th}$ site of a lane varies according to the occupancy state of its vertically opposite site (Fig. \ref{fig1}). Further, the particle can also hop vertically with a rate $w$. For simplicity, we consider the lane changing rate to be independent of interactions present in the model.

\begin{figure}
\includegraphics[trim = 60 600 80 40,clip,width=10cm,height=4.1cm]{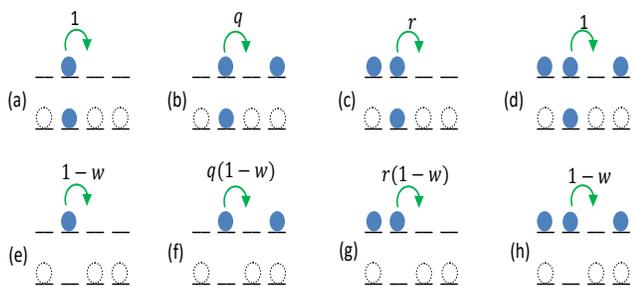}
\caption{Eight possible configurations of four consecutive vertical clusters in the bulk participating in the calculation of particle bulk currents.}\label{fig2}
\end{figure}

The dynamical rules of the model are as follows: for each time step, a lattice site $(i,l)$ is randomly selected from the two-channel system and any kind of transition is possible only when the target site is vacant. Random sequential update rules are adopted. In the bulk (see Fig. \ref{fig2}), a particle at $(i,l)^{th}$ site can hop to the empty site ($i+1,l)$ in eight different possible ways depending upon the occupancy state of sites $(i-1,l)$, $(i+2,l)$ and $(i,l')$, where ($l\neq l')$. These eight different ways can be understood in the following way. The occupancy states of sites $(i-1,l)$ and $(i+2,l)$ give rise to four different possibilities. Each of these four possibilities further splits into two sets depending upon the occupancy state of the site $(i,l')$.

The first possibility is that when there is neither deformation nor formation of the bond due to the particle movement (Fig. \ref{fig2}(a) and (e)). The second possibility is when formulation and deformation of bonds occur simultaneously (Fig. \ref{fig2}(d) and (h)). Furthermore, in both of these possibilities when there is a particle at site $(i,l')$, the rate of hopping is 1, alternatively in the absence of particle, the hopping rate is $1 - w$.
The third possibility is when the particle only breaks its bond from the left (Fig. \ref{fig2}(c) and (g)). The hopping rate in this case is $r$ when there is a particle at site $(i,l')$, otherwise it is $r(1-w)$. 
For the last possibility, $\tau_{i-1,l}$ = 0 and $\tau_{i+2,l}$ = 1 (Fig. \ref{fig2}(b) and (f)), the particle hops with a rate $q$ when $\tau_{i,l'} = 1$ otherwise the hopping rate is $q(1-w)$. In this case, there is only formation of the bond. Besides all these transitions, the particle at $(i,l)^{th}$ site can jump to site $(i,l')$ with a rate $w$ when the horizontal transition is not possible and the target site is empty.
In the proposed model, an open environment is considered, where particles enter from the left and leave the system from last sites. The effect of interactions at both the boundaries are considered as follows: At the left boundary, a particle can enter with rate $q\alpha$ if it forms the bond with its neighbor, otherwise the rate is $\alpha$ (Fig. \ref{fig1}). The transition rate of a particle at the right boundary depends not only on the occupancy state of its left neighbor but also on the state of $N^{th}$ site of the other lane. When the left neighboring site of ($N,l)^{th}$ site is empty, then for $\tau_{N,l'} = 1$ the particle leaves with a rate $\beta$ otherwise for $\tau_{N,l'} = 0$ the rate of leaving is $\beta(1-w)$. On the contrary, when $(N-1,l)$ is occupied, the rate of leaving of particle is $r\beta$ in the presence of particle at $(N,l')$ site, otherwise in the absence the rate is  $r\beta(1-w)$.

The formation and deformation of  bonds can be viewed as opposing chemical transitions \cite{teimouri2015theoretical} which give the following relationship between the transition
rates:
\begin{eqnarray}
\frac{q}{r} = e^{\lambda E} = \eta ~~~~\mbox{(say)}. \label{eq1}
\end{eqnarray}
Here $\lambda = (k_BT)^{-1} > 0 $ is a constant.
 Further, the hopping rates $q$ and $r$ can be explicitly expressed in terms of dimensionless parameter $\theta$ (0 $\leq \theta \leq 1)$ as
\begin{equation}\label{eq2}
q = e^{(\lambda\theta E)} = \eta^\theta, ~~~~~~~~r = e^{(\lambda (\theta - 1)E)} = \eta^{(\theta - 1)}.
\end{equation}
The splitting parameter ($\theta$) specifies the effect of energy on  these
transition rates.
When $\eta >1$ (or $E > 0$) the interactions are attractive and in this case the rate of formation of bond, $q$, is larger ($q \geq 1$) while the rate of deformation of bond, $r$, is smaller ($r \leq 1)$. But, the above rates for creating and breaking of bonds becomes smaller $(q \leq 1)$ and larger $(r \geq 1)$, respectively for repulsive interactions i.e. when $\eta < 1$ (or $E<0)$. In the absence of intra-channel interactions, the model reduces to symmetrically coupled original two-channel TASEP model \cite{pronina2004two}. Also, single channel TASEP model with nearest neighbor interactions is reproduced for the case of no inter-channel transitions \cite{teimouri2015theoretical,celis2015correlations}.

The proposed model is suitable to study the role of interactions in the collective dynamics on a symmetrically coupled two-channel transport process. The transition rules adopted here are consistent with the motion of motor proteins as the forward hoping rates in one channel depend not only on the number of bonds that remain unchanged, increased or decreased but also on the configuration of the other channel. The intra-channel interactions along with direct and indirect coupling in both the channels make the dynamics complex and difficult to analyze.
\subsection{Approximate Methods}
Single-channel original TASEP is one of the few models that has been solved by exact theoretical methods \cite{derrida1992exact,derrida1993exact,derrida1999bethe} under both open and closed boundary conditions.
For other variants of TASEP, collective dynamics have been analyzed by using  only approximate theoretical methods  \cite{juhasz2007weakly,popkov2004hydrodynamic,gupta2013coupling}. Also, for the case of an interactive single channel open TASEP system, recently, an approximation theory, called modified cluster mean field is proposed \cite{celis2015correlations}. However, for the case of multi-channel interactive TASEP system, neither an exact nor an approximate theory exist so far that can deal with the intra-channel interactions.
In this direction, we make a first attempt by developing an approximate theory that can handle the correlations produced in a coupled interactive two channel TASEP system.

In the following subsections, we first discuss an existing approximation approach that can incorporate the effect of inter-channel transitions and show that it is insufficient to incorporate intra-channel interactions. We then propose modified vertical cluster mean field (MVCMF) theory that incorporates the effect of both inter and intra-channel interactions.

\subsubsection{Vertical cluster mean field theory}
For a two-channel original TASEP system coupled either in symmetric, partially asymmetric or fully asymmetric fashion, vertical cluster mean field (VCMF) approach \cite{pronina2004two,gupta2013coupling} has been utilized to analyze steady state properties of the system and to produce density profiles which match exactly with those obtained with the direct Monte-Carlo simulation.


As the proposed model involves coupled two-channel TASEP system it will be an obvious choice to utilize the well examined VCMF approach for investigating the effect of interactions on the stationary properties of the system. For a non-zero vertical transition rate, VCMF theory deals with four possible vertical clusters of $i^{th}$ site of both the channels. Each single (one-site) vertical cluster for $i^{th}$ site can be found in one of the four states $\lbrace{0}\rbrace,\lbrace{1}\rbrace,\lbrace{2}\rbrace$ and $\lbrace{3}\rbrace$ with probabilities $V_0,V_1,V_2,$ and $V_3$, respectively, as shown in Fig. \ref{fig3}(a). Here, $V_0$ ($V_3$) denotes the probability when both of its sites are empty (occupied) and $V_1$ ($V_2$) represents the probability when only its upper (lower) site is occupied. The mutually exclusive and exhaustive nature of these probabilities implies
\begin{eqnarray}
\sum_{i}V_i = 1.\label{eq3}
\end{eqnarray}
In the stationary state, the bulk densities in each channel is computed as $\rho_1=V_{1} + V_{3}$, $\rho_2=V_2 + V_3$. The symmetric coupling in between the two channels provide $\rho_1=\rho_2=\rho$ \cite{pronina2004two}.

One can easily compute an expression of bulk current and check that the large interactions (i.e. $|E| \gg 1)$  cause bulk current to increase without any bound (see appendix A). Such behavior of bulk current is not physically acceptable. It is because, in the presence of stronger attractive interactions, particles will bound to form larger clusters, which hinders any movement of particles. While large repulsive interactions do not allow any two particles to bind together, thus causing current to approach zero.
Since, VCMF theory utilizes the simple mean field approximation which ignores the correlations between two neighboring vertical clusters, it fails to compute the appropriate bulk current of the system\cite{teimouri2015theoretical}. Thus, correlations affect the movement of particles and can not be ignored.
To overcome the incapability of the mean field approximation, we now develop a generalized theory called modified vertical cluster mean field theory (MVCMFT) for multi-lanes system that predicts appropriate finite maximal current in the presence of strong attractive interactions and genuinely produces the steady state phase diagrams for the two-channel system.

\subsubsection{Modified Vertical Cluster Mean Field (MVCMF) Theory}
\begin{figure}
\begin{center}
\includegraphics[width=8.8cm,height=3.7cm]{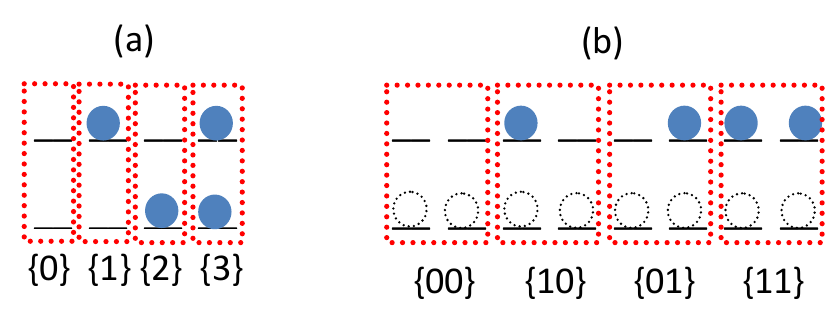}
\caption{Four different states for (a) one-site vertical cluster (b) two-site vertical cluster. Here, a dotted-open circle denotes that the site is either empty or occupied. A filled circle indicates the occupied site whereas absence of circle means the empty site.}\label{fig3}
\end{center}
\end{figure}
In the approach, we examine the role of some correlations by considering two neighboring vertical clusters. Based on the configuration of two neighboring sites of a lane, each two-site (two neighboring) vertical cluster is classified in four different states $\lbrace00\rbrace,\lbrace{10}\rbrace,\lbrace{01}\rbrace$, and $\lbrace{11}\rbrace$ with probabilities $H_{00},H_{10},H_{01}$, and $H_{11}$, respectively, as shown in Fig. \ref{fig3}(b). The normalization condition for these probabilities leads to
\begin{eqnarray}
\sum_{i,j=\lbrace{0,1}\rbrace} H_{ij} = 1.\label{eq4}
\end{eqnarray}
At the steady state, for the translational invariant system, the Kolmogorov consistency conditions provide the following relationship for uniform bulk density $\rho$
\begin{eqnarray}
H_{10} + H_{11} = \rho ~,~~~~~~~~~~H_{01} + H_{11} = \rho. \label{eq5}
\end{eqnarray}
To incorporate the effect of inter as well as intra-channel interactions, we compute the probabilities of each two-site vertical clusters, defined in Fig. \ref{fig3}(b), explicitly in terms of one-site vertical cluster as
\begin{eqnarray}
H_{10} &=& \frac{(V_1 + V_3)(V_0 + V_2)}{V_0 + V_2 + \eta(V_1 + V_3)},  \label{eq.(6a)} \\
H_{11} &=& (V_1 + V_3) - H_{10},  \label{eq.(6b)}\\
H_{00} &=& (V_0 + V_2) - H_{10}, \label{eq.(6c)}\\
\hspace{-2.5cm}\mbox{and}~~~~~~~~~~~~~H_{01} &=& H_{10}.\label{eq.(6d)}
\end{eqnarray}
Clearly, equations (\ref{eq.(6a)}) - (\ref{eq.(6d)}) respect the normality condition given by Eq. (\ref{eq4}).
 To understand the approximation of $H_{10}$ in Eq. (\ref{eq.(6a)}), we first discuss the case when there are no intra-channel interactions i.e. $\eta$ = 1. Using mean field approximation, the probability $H_{10}$ can be written as a product of the probabilities of its each one-site vertical cluster. First vertical cluster of configuration $\lbrace{10}\rbrace$ can exists in any of the two states $\lbrace 1 \rbrace$ and $\lbrace 3 \rbrace$ with probability ($V_1$ + $V_3$). While the second vertical cluster can exist in one of the two states $\lbrace 0 \rbrace$ and $\lbrace 2 \rbrace$ and thus its probability is ($V_0$ + $V_2$). This implies $H_{10} = (V_1 + V_3)(V_0 + V_2)=\rho(1-\rho)$. However, the probability $H_{10}$ in Eq. (\ref{eq.(6a)}) includes the effect of interactions between two occupied vertical clusters. It is the product of two probabilities. The first one is $(V_1 + V_3)$, the probability for first vertical cluster of configuration $\lbrace 10 \rbrace$ to either be in state $\lbrace 1 \rbrace$ or $\lbrace 3 \rbrace$. The remaining term, $\frac{(V_0 + V_2)}{(V_0 + V_2) + \eta(V_1 + V_3)}$, is the conditional probability of the second vertical cluster to be in state $\lbrace 0 \rbrace$ or $\lbrace 2 \rbrace$ knowing that the first vertical cluster is in state $\lbrace 1 \rbrace$ or $\lbrace 3 \rbrace$. Note that when second vertical cluster is occupied, its particles can interact with the particles of the first vertical cluster; this effect is incorporated by the factor $\eta$ appearing in the denominator. Other equations (\ref{eq.(6b)}) - (\ref{eq.(6d)})  can easily be computed using Eq. (\ref{eq5}). All these equations hold equally well for the limiting cases.
 In the absence of intra-channel interactions ($\eta = 1$), $H_{10} = (V_0 + V_2)(V_1 + V_3) = \rho(1 - \rho)$. For very strong repulsions ($\eta \rightarrow 0$), $H_{10} = V_1 + V_3 = \rho$ which is same as for the motion of non-interacting dimers on the lattice \cite{lakatos2003totally}. Under very large attractive strength ($\eta \rightarrow \infty)$ $H_{10} \rightarrow 0$, which seems to be justified, as the whole system is expected to be fully occupied without any vacancies.
 

  We now compute the total particle flux in the system.
According to the system dynamics,
there are total eight configurations in the bulk that participate in the computation of particle current (flux) per channel as illustrated in Fig. \ref{fig2}. The total flux per channel can be written as
\begin{equation}
J_{bulk} = \sum_{i = a}^{h}J_i, \label{eq7}
\end{equation}
where $J_a$, $J_b$, $\cdots$, $J_h$ denotes particle flux corresponding to the eight different configurations in Fig. \ref{fig2}(a-h), respectively. Each of these configuration consists of four consecutive one-site vertical cluster. In all these cases, the current is measured only when there is the particle movement from the upper site of second vertical cluster to the upper site of third vertical cluster. The movement depends on the occupancy states of first, fourth and the lower site of second vertical cluster.\\
We first approximate particle current corresponding to the configurations whose lower site of the second vertical cluster is filled (Fig. \ref{fig2}(a)-(d)).
For configuration (a), $J_a$ is expressed as
\begin{equation}
J_a = \gamma S_1 \Bigg(\frac{H_{00}}{H_{00} + V_1+V_3}\Bigg).\label{eq8}
\end{equation}
The above expression can be understood as the product of three terms. The first factor $ \Big[ \gamma(\gamma + (1-\gamma)) \Big] $ gives the
 probability of the first vertical cluster. $\gamma = (1+\eta)^{-1}$ is a Boltzmann's factor and here it provides the probability of the upper site of first vertical cluster in occupied state. $[\gamma + (1-\gamma)]$ is the probability for its lower site which can be either empty or occupied.
The second factor, $S_1$, is an approximated probability of the two-site vertical cluster whose first vertical cluster is in state $\lbrace3\rbrace$ and second vertical cluster is either in state $\lbrace0\rbrace$ or $\lbrace2\rbrace$ (Fig. \ref{fig2} (a)) and is computed as
\begin{equation}
S_1 = \frac{V_3(V_0 + V_2)}{(V_0 + V_2) + \eta ( V_1 + V_3)}.\label{eq9}
\end{equation}
Here, the factor $\eta$ in the denominator represents the interaction between particles of first and second vertical clusters. Finally,
the last and the third factor of $J_{a}$
 indicates the probability of the fourth vertical cluster to be in state $\lbrace{0\rbrace}$ or $\lbrace{2}\rbrace$ i.e. its upper site is empty and the lower site is either empty or occupied.
It is the normalization of the two probabilities corresponding to the two possible configurations in which the fourth vertical cluster can exist.
The two configurations are:\\
(i) When the upper site is occupied
  with probability
  $ V_1 + V_3 = \rho$\\
(ii) When the fourth vertical cluster is either in state $\lbrace{0\rbrace}$ or $\lbrace{2}\rbrace$. Since, the third vertical cluster is in one of the states $\lbrace{0\rbrace}$ or $\lbrace{2}\rbrace$, the probability for the fourth vertical cluster, in this case, is given by $H_{00}$.\\
So, the normalized probability of the fourth vertical cluster is $\Bigg(\frac{H_{00}}{H_{00}+(V_1+V_3)}\Bigg)$.

By similar arguments, particle currents for configurations (b) - (d) can be computed as:
\begin{eqnarray}
J_{b} &=& q\gamma S_1\Bigg(\frac{V_1 + V_3}{H_{00}+V_1+V_3}\Bigg),\label{eq10}\\
J_{c} &=& r(1-\gamma) S_1\Bigg(\frac{H_{00}}{H_{00}+V_1+V_3}\Bigg),\label{eq11}\\
\mbox{and}~~~J_{d} &=& (1-\gamma)S_1\Bigg(\frac{V_1 + V_3}{H_{00}+V_1+V_3}\Bigg).\label{eq12}
\end{eqnarray}

Now we calculate flux for configurations in Fig. \ref{fig2}(e-h), which are similar to configurations (a) - (d) except that the lane changing of the particle is allowed. Thus flux corresponding to the configurations (e) - (h) can be approximated as
\begin{eqnarray}
J_{e} &=& (1-w)\gamma S_2\Bigg(\frac{H_{00}}{H_{00}+V_1+V_3}\Bigg),\label{eq13}\\
J_{f} &=& q(1-w)\gamma S_2\Bigg(\frac{V_1 + V_3}{H_{00}+V_1+V_3}\Bigg),\label{eq14}\\
J_{g} &=& r(1-w)(1-\gamma)S_2\Bigg(\frac{H_{00}}{H_{00}+V_1+V_3}\Bigg),\label{eq15}\\
\mbox{and}~~J_{h} &=& (1-w)(1-\gamma)S_2\Bigg(\frac{V_1 + V_3}{H_{00}+V_1+V_3}\Bigg).\label{eq16}
\end{eqnarray}
Here, the factor $S_2$ represents the probability of the two-site vertical cluster whose first cluster is in state $\lbrace{1}\rbrace$ and second cluster is either in state $\lbrace{0}\rbrace$ or $\lbrace{2}\rbrace$ and is approximated as
\begin{equation}
S_2 = \frac{V_1(V_0 + V_2)}{V_0 + V_2 + \eta( V_1 + V_3)}.\label{eq17}
\end{equation}
Note that $S_1 + S_2 = H_{10}$.\\
Utilizing Eq. (\ref{eq8}) and Eqs.(\ref{eq10}-\ref{eq16}), in Eq. (\ref{eq7}), the total bulk current is obtained as
\begin{equation}
J_{bulk} = \frac{AH_{00} + B\rho}{H_{00} + \rho}\Bigg(S_1 + (1-w)S_2\Bigg).\label{eq18}
\end{equation}
where auxiliary functions $A$ and $B$ are defined as
\begin{equation}
A = \frac{1+r\eta}{1+\eta},~~~~~~~~~~~~~~  B =  \frac{q+\eta}{1+\eta}.\label{eq19}
\end{equation}
Since, $J_{bulk}$ involves $S_1$, $S_2$, $H_{00}$ and $\rho$ that are in terms of one-site vertical cluster  probabilities, $J_{bulk}$ is also in terms of $V_i^{'}s$.
To write $J_{bulk}$ explicitly as a function of single ordered parameter $V_3$, we solve the master equation for these one-site vertical cluster probabilities in the mean field approximation at steady state, to get,\
\begin{eqnarray}
\begin{aligned}
&V_3V_0 = (1-w)V_1^2 \label{eq20},\\
\hspace{-4cm}\mbox{and}~~~~~~~~&V_1 = V_2.\label{eq21}
\end{aligned}
\end{eqnarray}
The normalization condition from Eq. (\ref{eq3}) and Eq.\ref{eq20} gives $V_0 = 1-2V_1-V_3$, which further implies $V_1$ as
\begin{equation}
V_1 =
\hspace{0.2cm}
  \left\{
    \begin{array}{l}
      \frac{-V_3 + \sqrt{V_3^2 + (1-w)V_3(1-V_3)}}{1-w},~~~~~ (w \neq 1)\\
      \frac{1-V_3}{2},~~~~~~~~~~~~~~~~~~~~~~~~~~~~~~(w = 1).
    \end{array}
  \right.\label{eq22}
\end{equation}


All dynamic properties of the system can now be calculated using $J_{bulk}(V_3)$ for any given value of interaction energy $E$, splitting parameter $\theta$ and coupling rate $w$.
The expression for $J_{bulk}$ holds equally well for all the limiting cases.
For the case of no interactions  $(\eta = 1)$, $J_{bulk}$= $(V_3 + (1-w)V_1)(1-V_1-V_3)$, which is same as reported in Ref. \cite{pronina2004two}.
For very large repulsive interactions $(\eta \rightarrow 0)$,
\begin{equation}
J_{bulk} \rightarrow \frac{(V_3 + (1-w)V_1)(1-2(V_1 + V_3))}{(1-V_1-V_3)}. \label{eq22a}
\end{equation}
This result is reasonable as for a special case $w$ = 0 $J_{bulk}$ reduces to $\frac{\rho(1-2\rho)}{(1-\rho)}$ \cite{celis2015correlations}.
Under very large attractions $(\eta \rightarrow \infty)$, $J_{bulk} \rightarrow 0$. Such behavior of bulk current matches the expectation that large attractions force particles to form longer clusters which hinder their movement. It also justifies that our proposed generalised approach has overcome the pitfall of the VCMF approach.

Now, we discuss the current at the two boundaries where dynamics are totally governed by the exit and entrance rates. These currents are computed in the similar fashion used for computing bulk current and are expressed as
\begin{eqnarray}
J_{entr} &= \frac{\alpha(1-V_1-V_3)[1 - (2 - \eta + q)(V_1 + V_3)]}{1 - (1 - \eta)(V_1 + V_3)},\label{eq23}
\end{eqnarray}
and
\begin{eqnarray}
J_{exit} &= \frac{\beta((1-w)V_1 + V_3)[1 - (1 - r\eta)(V_1 + V_3)]}{1 - (1 -\eta)(V_1 + V_3)}.\label{eq24}
\end{eqnarray}
For $\eta = 1$ i.e. no interactions, $J_{entr} = \alpha(1-V_1-V_3) = \alpha(1 - \rho) $ and $J_{exit} = \beta((1-w)V_1 +V_3)$, which is likely. When $\eta \rightarrow 0$ i.e. in the presence of strong repulsive interaction $J_{entr} = \alpha(1 - 2(V_1 + V_3)) =  \alpha(1-2\rho)$ and $J_{exit} = \beta((1-w)V_1 + V_3)$.
 For large attractive interactions i.e. $\eta \rightarrow \infty$, $J_{entr} = J_{exit} = 0$, which is as expected.
\textbf{\subsection{Phase diagrams}}

To analyze the effect of interactions on symmetrically coupled TASEP system we construct stationary density profiles and phase diagrams in the parameter space $(\alpha, \beta)$ based on the theoretical investigation presented in the previous section.
In the absence of interactions, the model under study reduces to the original case of symmetrically coupled TASEP with three dynamical phases Low-density (LD), High-Density (HD) and Maximal-Current (MC) \cite{pronina2004two}.
It has been observed that the consideration of interactions on a single-channel original TASEP system did not alter the topology of the phase diagram but shifted the phase boundaries \cite{celis2015correlations}.
It is reasonable to expect that although the presence of interaction in symmetrically coupled system will preserve the nature of the density profiles and the qualitative properties of the phase diagram but the location of triple points, phase boundaries, values of stationary maximal bulk currents and densities will change.

We now discuss the properties of the three different stationary phases and provide the explicit expression for computing the phase boundaries separating them.

 \textbf{\textit{Entrance Dominated phase (LD)}}: The system dynamics in this phase are solely governed by the entrance rate. As a result, the bulk current of the system matches with the entrance current. This continuity of the stationary current provides a relationship between the entrance rate $\alpha$ and the order parameter $V_3$ from the following equation:
 \begin{eqnarray}
 \begin{aligned}
\alpha &= (V_3+(1-w)V_1) \\ 
&\times \Bigg[
 \frac{A(1 - \rho)(1 -\rho(2 - \eta))+B\rho(1 - \rho(1 - \eta))}{(1 + \rho(-2+\eta+q))((1-\rho)^2 + \rho \eta)}\Bigg].\label{eq25}\end{aligned}
\end{eqnarray}
After substituting $\rho_{LD}$ = $V_{3}^{LD} + V_{1}^{LD}$ and $V_{1}^{LD}$ from Eq. (\ref{eq22}), the above nonlinear equation can be solved for a relevant root $V_{3}^{LD}$. This subsequently provides particle bulk density in the LD phase in terms of $\alpha$ using the relation $\rho_{LD}$ = $V_{3}^{LD} + V_{1}^{LD}$ and Eq. (\ref{eq22}). Further, bulk current can be computed from Eq. (\ref{eq18}) for any value of coupling rate $w$.

We check the validity of the above estimates for special cases. When particles do not interact with each other ($\eta = 1$), Eq. (\ref{eq25}) simplifies to $\alpha$ = $\sqrt{V_3(1-w(1-V_3))}$ and the density function reduces to $\rho_{LD}$ = $\frac{2\alpha + 1 - w -\sqrt{(1-w)^2 + 4w\alpha^2}}{2(1-w)}$, which agrees well with the known results of two-channel symmetrically coupled original TASEP system~\cite{pronina2004two}. In the limit of infinite repulsive interactions ($\eta \rightarrow 0$), Eq. (\ref{eq25}) yields
\begin{eqnarray}
\alpha = \frac{(1-w)\sqrt{V_3(1-w(1-V_3))}}{1-w(1-V_3)-\sqrt{V_3(1-w(1-V_3))}}.\nonumber
\end{eqnarray}
One can solve above equation for $V_3^{LD}$ and thus can obtain $\rho_{LD}$ = $\frac{\alpha(1+\alpha)}{(1+\alpha)^2 - w}$. Furthermore, for a special choice of $w$ = 0, $\rho_{LD}$ becomes $\left( \frac{\alpha}{1 + \alpha}\right) $ \cite{lakatos2003totally}.
For the limiting case of infinite attractive interactions ($\eta \rightarrow \infty$), bulk current tends to zero which is possible only for $\alpha=0$. Thus, the continuity of the entrance and bulk current implies that LD phase does not exist under very large attractive interactions.\\
\textbf{\textit{Exit Dominated Phase (HD)}}: System dynamics in this phase are dominated by the exit rate. The condition $J_{bulk} = J_{exit}$, yields a coupled relationship between order parameter and exit rate from the following relationship:
\begin{equation}
\beta = \frac{(1 - \rho)\Big[A(1 - \rho)(1 - \rho(2 - \eta)) + B\rho(1 -\rho(1-\eta) )\Big]}{(1 + \rho (r \eta - 1))((1-\rho)^2 + \rho \eta)}.\label{eq26}
\end{equation}
From the above expression and utilizing relation $\rho_{HD} = V_3^{HD} + V_1^{HD}$ and Eq. (\ref{eq22}), the density of full vertical cluster in HD phase, $V_3^{HD}$, is calculated as a function of $\beta$. Subsequently, particle density, $\rho_{HD}$, and current, $J_{HD}$, is obtained from relation $\rho_{HD} = V_3^{HD} + V_1^{HD}$ and Eq.  (\ref{eq24}), respectively, as a function of $\beta$.
For the case of no interaction ($\eta=1$), Eq. (\ref{eq26}) simplifies to
\begin{eqnarray}
\beta = \frac{1-w(1-V_3) - \sqrt{V_3(1-w(1-V_3))}}{1-w},\nonumber
\end{eqnarray}
 which further provides $\rho_{HD} = 1 - \beta$, that is exactly same as obtained in Ref. \cite{pronina2004two}.
Under the case of very large repulsions ($\eta \rightarrow 0$), Eq. (\ref{eq26}) reduces to
\begin{eqnarray}
\beta = \frac{1 - w(1-2V_3) - 2\sqrt{V_3(1-w(1-V_3))}}{1 - w(1-V_3) - \sqrt{V_3(1-w(1-V_3))}}, \nonumber
\end{eqnarray}
that gives the density of fully filled vertical cluster in the high density phase explicitly in terms of exit rate as
$$V_3^{HD} = \frac{\beta(2w-1) + \sqrt{(2-\beta)^2 -4w(1-\beta)} -2(1-w)}{2w(2-\beta)} .$$
From the above expression, the density $\rho_{HD}$  in terms of coupling rate $w$ can be obtained from relation $\rho_{HD} = V_3^{HD} + V_1^{HD}$ and Eq. (\ref{eq22}). Particularly, for $w = 0$, $\rho_{HD}$ = $\frac{1-\beta}{2-\beta}$, which matches with the case of non-interacting dimers \cite{lakatos2003totally}.
Lastly for infinite attractions ($\eta \rightarrow \infty$), the bulk current and thus entry and exit current tend to zero for all values of $\beta$ leading to a fully occupied system.\\

\begin{figure*}
\begin{center}
\includegraphics[trim = 0 30 38 00,clip,width=8cm,height=7cm]{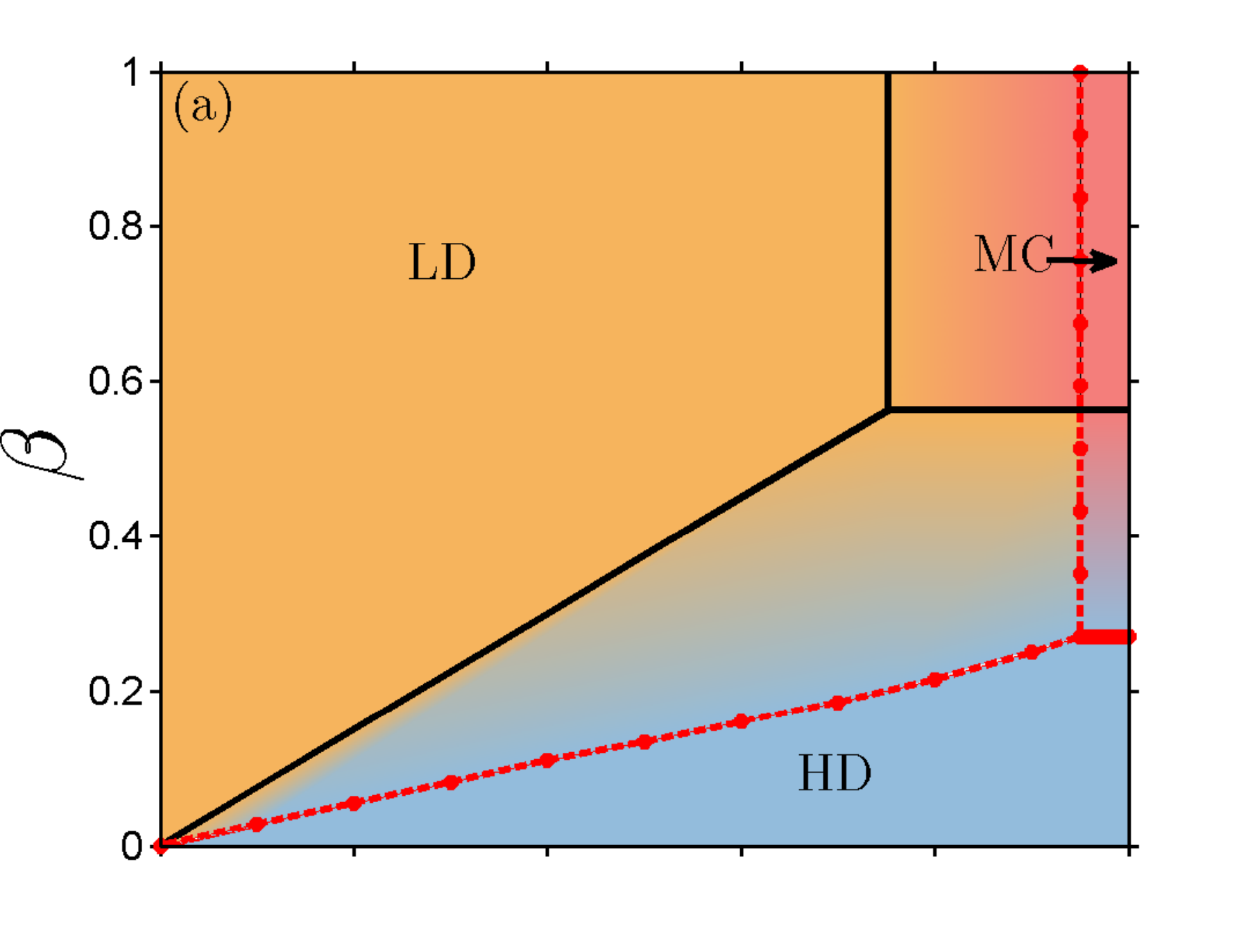}
\includegraphics[trim = 60 30 20 0,clip,width=7.7cm,height=7cm]{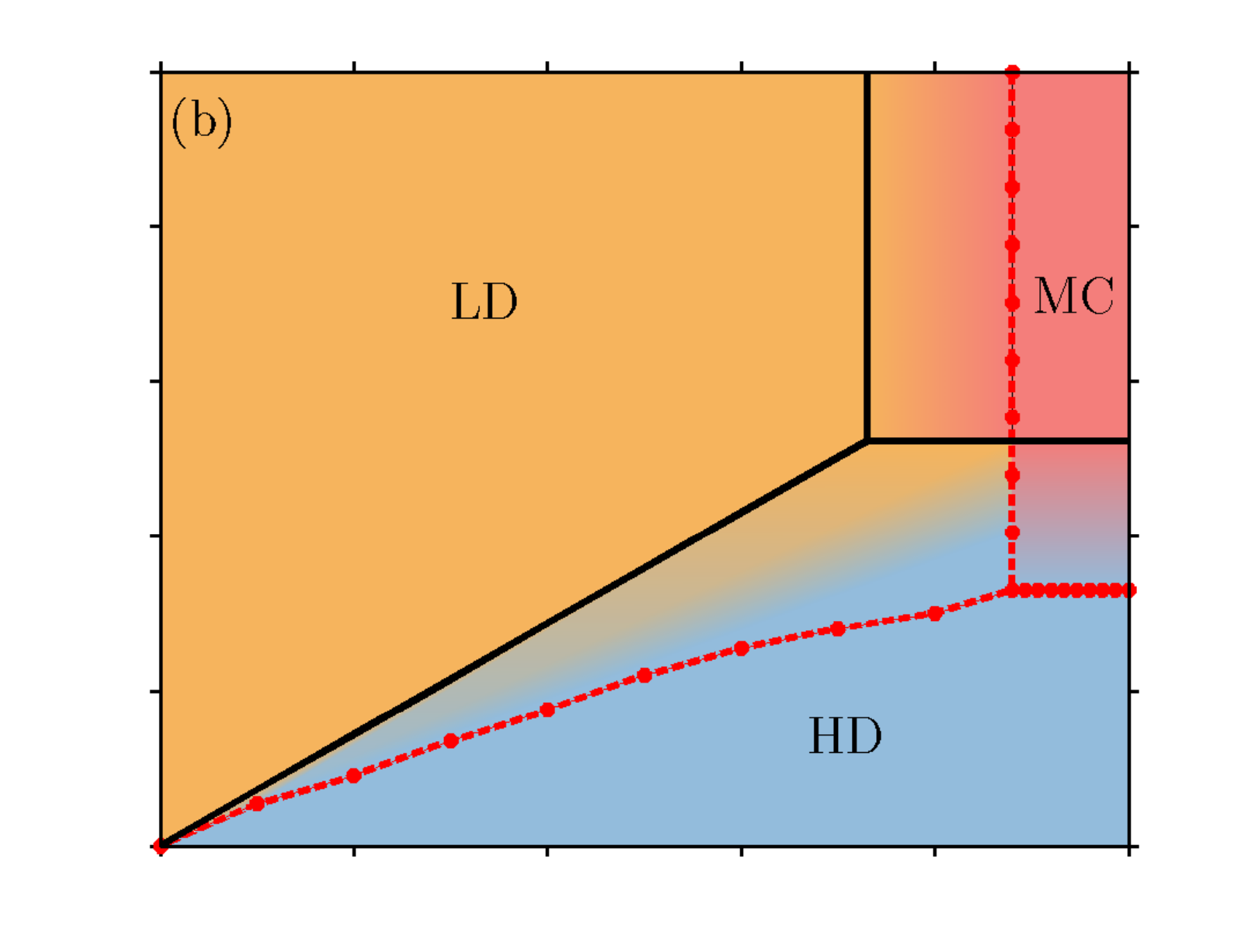}
\includegraphics[trim = 0 0 40 20,clip,width=7.9cm,height=7cm]{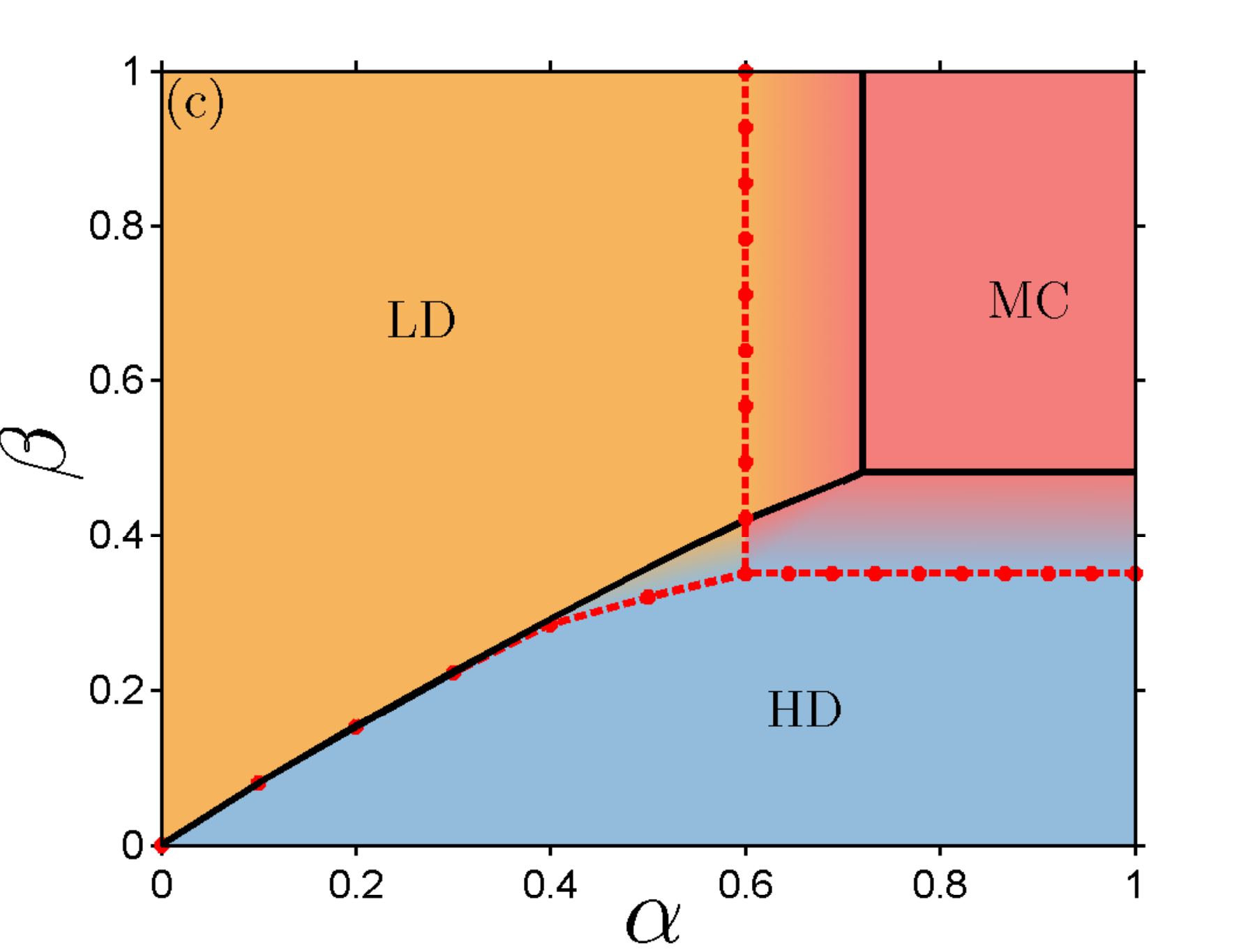}
\includegraphics[trim = 60 00 30 20,clip,width=7.53cm,height=7cm]{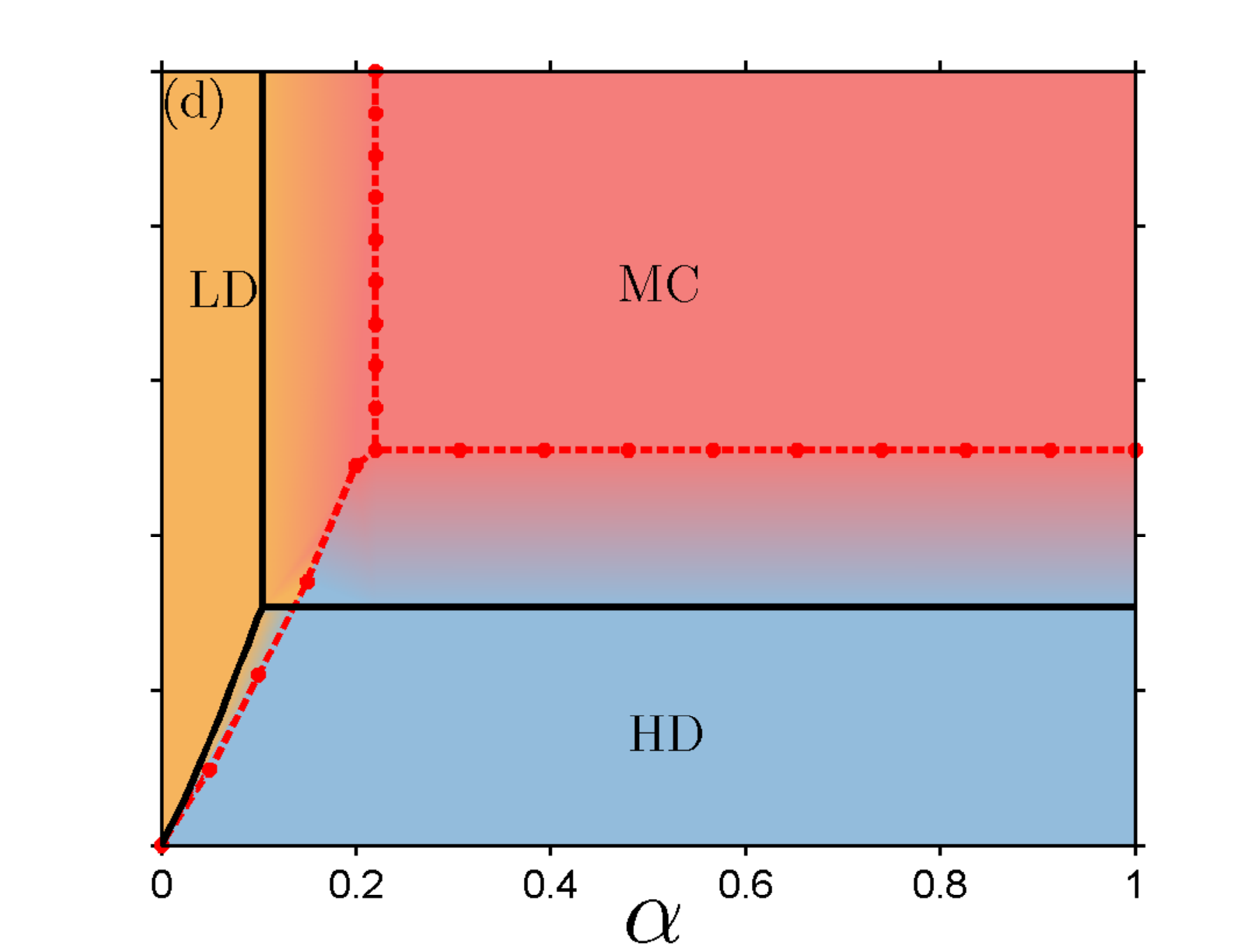}
\caption{Stationary phase diagrams for interactive two-channel symmetrically coupled TASEP with coupling strength $w = 0.2$ for different interaction strengths and splitting parameter: (a) $E = -1.6\lambda^{-1}$, $\theta$ = 0.25; (b) $E = -1.2\lambda^{-1}$, $\theta$ = 0.5; (c) $E = -0.7\lambda^{-1}$, $\theta$ = 0.75; (d) $E =1.6\lambda^{-1}$, $\theta$ = 0.5. Solid lines and dotted lines with marker indicate theoretical and simulations results, respectively.}\label{fig4}
\end{center}
\end{figure*}

\textbf{\textit{Maximal Current Phase (MC)}}: In this phase, the bulk current is dominated by the bulk processes and remains unaffected by the boundary rates. For current $J_{bulk}$ to be maximum, the following condition:
\begin{equation}
 \frac{\partial J_{bulk}}{\partial V_3} =0,\label{eq27}
\end{equation}
must be satisfied. It is worth to mention that there is no complexity involved in obtaining the general analytic expression of the above equation. However, the complexed form is too lengthy to mention here. The physically relevant root, $V_3^{MC}$, can be determined by numerically solving Eq. (\ref{eq27}) for any general set of constants $E$, $\theta$ and $w$.
Thus, the stationary properties namely density, $\rho_{MC}$, and maximal current, $J_{MC}$, can be calculated by using relation $\rho_{MC} = V_3^{MC} + V_1^{MC}$ and Eq. (\ref{eq18}), respectively, where $V_1^{MC}$ is obtained from Eq. (\ref{eq22}).

We now test the validity of our analytical results for some special cases.
For the case of no interactions, the lengthy expression obtained from the general Eq. (\ref{eq27}) simplifies to
\begin{eqnarray}
\begin{aligned}
4w^2V_3^2 &- \Big[4w\sqrt{V_3(1-w(1-V_3))} -5w(1-w)\Big]V_3  \nonumber\\ &-2\sqrt{V_3(1-w(1-V_3))}(1-w)+(1-w)^2=0.\end{aligned}
\end{eqnarray}
 The above equation matches with the known result for the original two-channel TASEP model with symmetric coupling and can be solved for relevant root, $V_3^{MC}$, for any value of $w$ \cite{pronina2004two}.
For the case of large repulsion ($\eta \rightarrow 0$), Eq. (\ref{eq27}) yields,
 \begin{eqnarray}
 \begin{aligned}
 &4w(1+w)V_3^2 + (-6w^2 + w(4 - 8\sqrt{V_3(1+(V_3-1)w)})
  \nonumber \\ &+ 2)V_3 + (1-w)(1-w-4\sqrt{V_3(1+(V_3-1)w}) = 0.\end{aligned}
 \end{eqnarray}
The physically reasonable root for above equation, $V_3^{MC}$, which gives nonzero flux in the system, can be found out for any value of $w$. In particular, for $w = 0$, one can get $V_3^{MC}$ = $\frac{1}{2}(3 - 2\sqrt{2)}$, $\rho_{MC} = 1 - \frac{1}{\sqrt{2}}$ and $J_{bulk} \simeq 0.1716$, which matches for the case of one channel interacting TASEP model in the limit of large repulsion \cite{celis2015correlations}.\\
We now investigate three different two-phase coexistence lines.

\textbf{(i) LD-MC}:
As the transition from LD to MC phase is second order in nature, the bulk densities in both the phases will be equal at the transition line. Since, $\rho_{LD}$ depends on $\alpha$ and an increment in $\alpha$ causes rise in the bulk density in LD phase; there must exist a critical value of $\alpha$, say $\alpha_c$, at which $\rho_{LD}$ = $\rho_{MC}$. The line $\alpha$ = $\alpha_c$ gives the phase boundary line for these two phases.


  \textbf{(ii) HD-MC}: The phase coexistence line, HD-MC, can similarly be obtained by using the condition of continuous transition of bulk density from HD to MC line. In general, $\rho_{HD}>\rho_{MC}$ and $\rho_{HD}$ decreases continuously with increase in $\beta$. At a critical value of $\beta$ = $\beta_c$ (say), $\rho_{HD}$ = $\rho_{MC}$. The line $\beta$ = $\beta_c$ is the required phase boundary line separating these two phases.

  \textbf{(iii) LD-HD}: The transition from LD to HD phase is first order discontinuous and the phase transition line can be determined from continuity condition for current i.e. $J_{entr}$ = $J_{exit}$, that provides the following relationship between $\alpha$ and $\beta$.  
\begin{eqnarray}
\begin{aligned}
\frac{\beta}{\alpha} &= \left[  \frac{(1-\rho_{LD})(1 - 2\rho_{LD} + q\rho_{LD} + \eta \rho_{LD}}{(V_{3}^{HD} + (1-w)\rho_{HD}-V_{3}^{HD})(1-\rho_{HD}(1 - r\eta)}\right]\\
& \times \left[ \frac{(1 - \rho_{HD} + \eta \rho_{HD})}{(1 - \rho_{LD} + \eta \rho_{LD})}\right].
 \end{aligned}
\end{eqnarray}
 Here, $V_{3}^{LD}$, $\rho_{LD}$ and  $V_{3}^{HD}$, $\rho_{HD}$  are densities obtained by solving Eqs. (\ref{eq25}) and (\ref{eq26}), respectively. Starting from the origin, the transition line ends meeting up the triple point $(\alpha_c,\beta_c)$, where all the three phase boundaries intersect. At this point, $\rho_{LD} = \rho_{HD} = \rho_{MC}$.

 \section{Results and Discussions}
 In order to examine the effect of attractive as well as repulsive interactions on the collective dynamics of motor proteins working on two-channel coupled system and to test the applicability of the proposed approximate theoretical framework, we plot phase diagrams obtained theoretically as well as by Monte-Carlo simulations for different repulsive and attractive strength of interactions under a fixed coupling rate $w$ = 0.2 in Fig. \ref{fig4}.
It is worth to mention here that as our generalized approach agrees with the known vertical cluster mean field approach for the case of no interactions, the theoretical and simulation results of the phase diagrams matches exactly.

  On comparing the phase diagrams for different interaction strength and splitting parameter (Fig. \ref{fig4}), it is evident that theoretical results agree quite well with the simulation results for both relatively weak attractive and repulsive interaction. For stronger interactions, the theoretical and simulation results agree mostly at the qualitative level.
Additionally, for a fixed coupling strength, the LD-MC and LD-HD phase transition lines shift rightwards and downwards, respectively, with an increase in the strength of repulsive interaction which further causes the enlargement of LD region in the phase plane. While for attractive interaction, LD (HD) phase shrinks (enlarges) due to the movement of LD-MC and HD-MC line towards the left and up, respectively.  The reason behind such behavior is understood as follows: Repulsive interaction reduces the effective entrance rate of the system as well as drive the particles away from each other which opposes the existence of large particle cluster. Both situations favor the LD phase. While attractions increase the effective entrance rate and decrease the exit rate, both conditions force particles to form bigger clusters leading to high density.

\begin{figure}
\begin{center}
\includegraphics[width = 8.6cm, height =6.6cm]{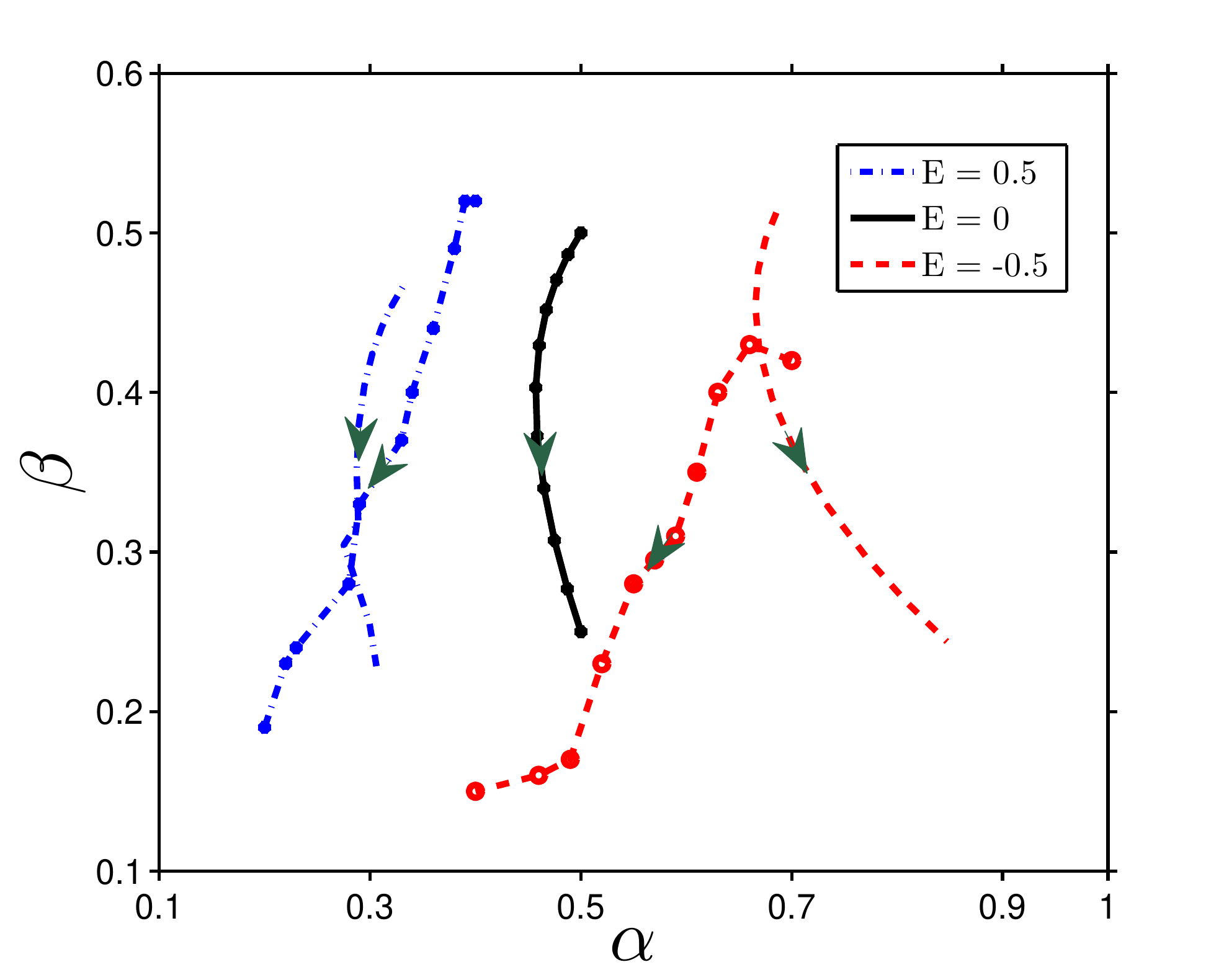}
\caption{Triple points as a function of coupling rate $w$ under weak attractive and repulsive interactions split symmetrically. In the case of interactions, different dotted lines with and without marker represents simulations and theoretical results, respectively. Solid line with and without marker highlights the case of no interactions. The pointing arrows indicate $w$ increasing from 0 to 1.}\label{fig5}
\end{center}
\end{figure}

\begin{figure*}
\begin{center}
\includegraphics[trim = 0 0 20 00,clip,width=3.45cm,height=3.4cm]{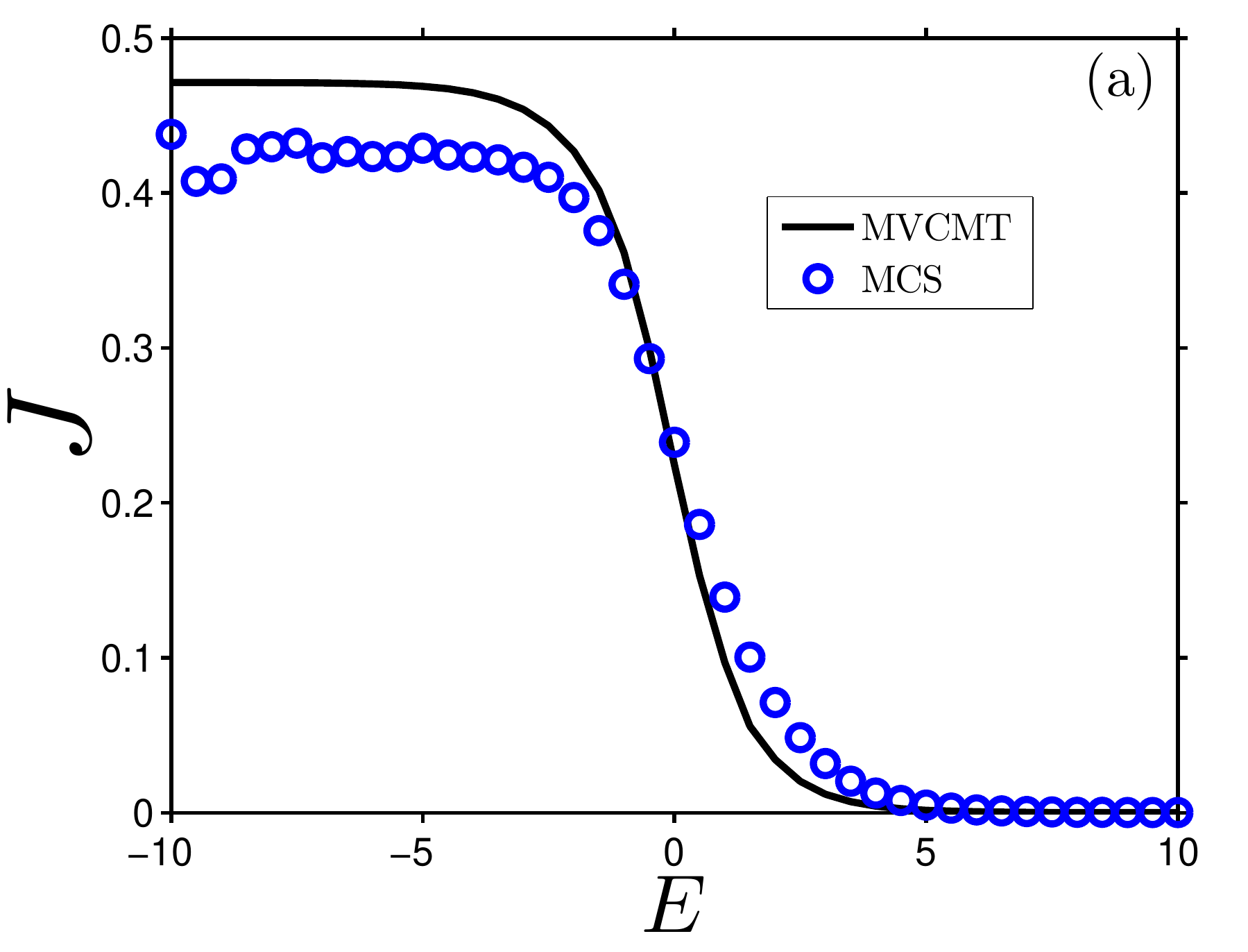}
\includegraphics[trim = 45 0 47 0,clip,width=3.4cm,height=3.4cm]{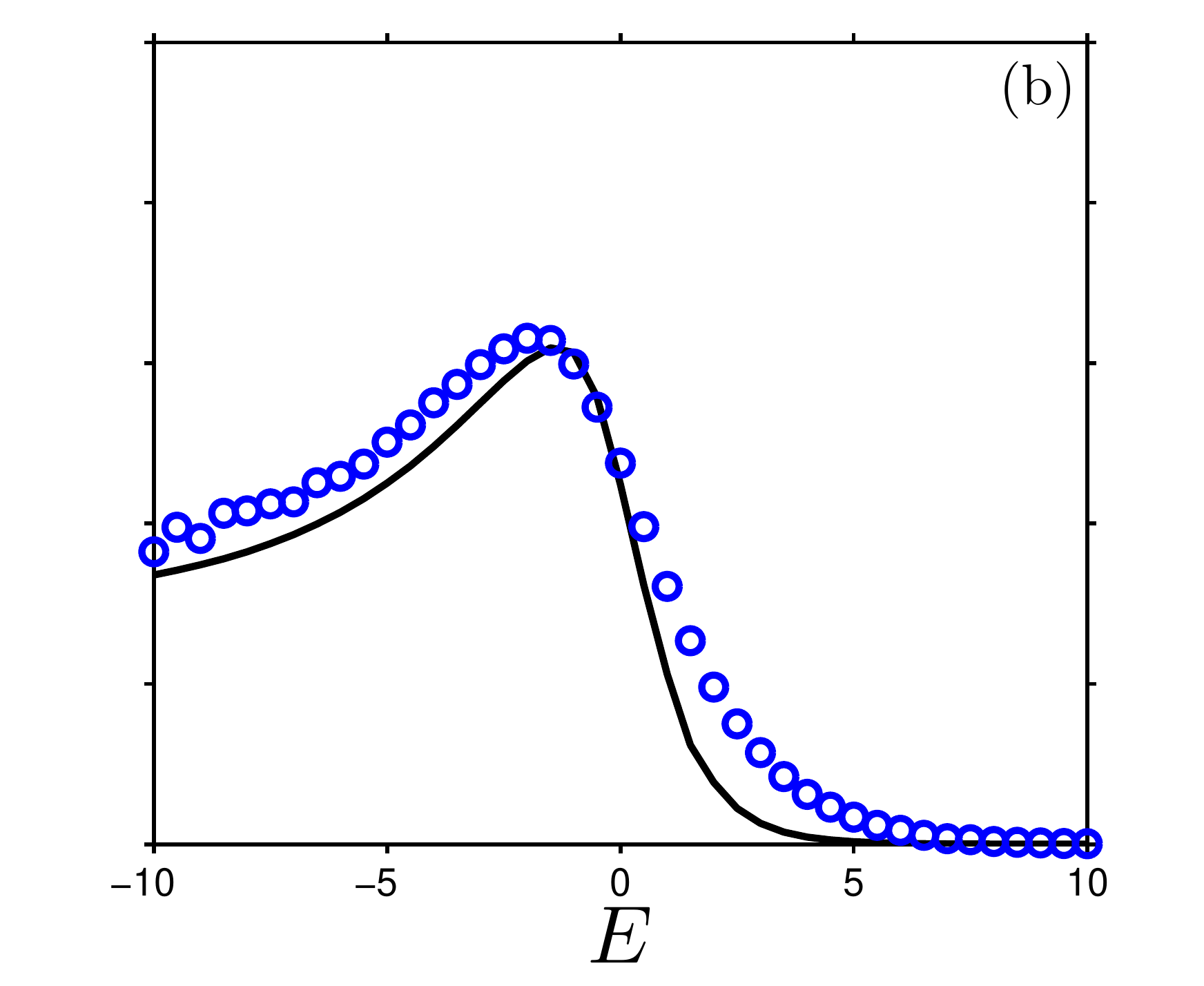}
\includegraphics[trim = 47 3 35 0,clip,width=3.4cm,height=3.385cm]{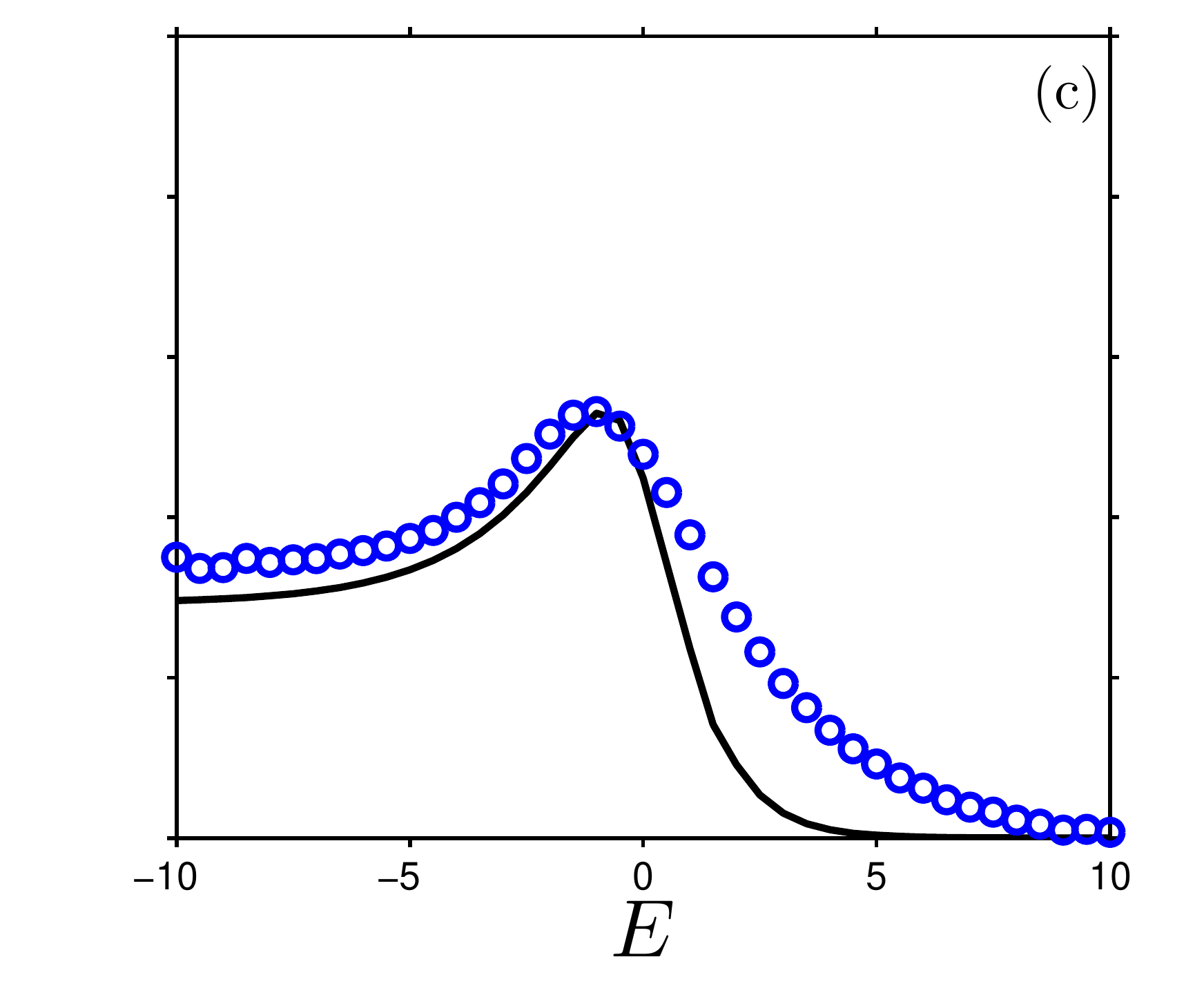}
\includegraphics[trim = 37 0 22 0,clip,width=3.5cm,height=3.4cm]{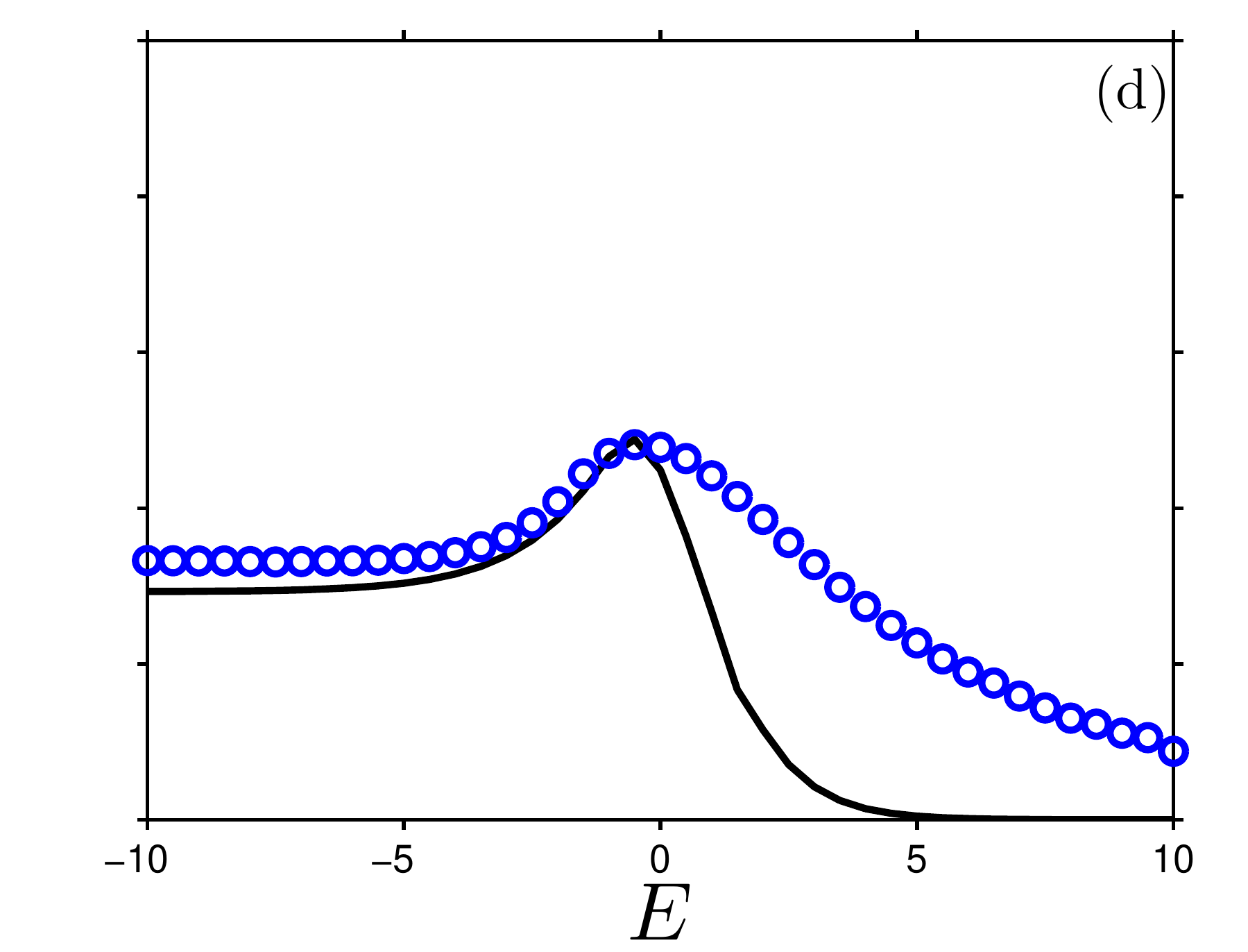}
\includegraphics[trim = 47 0 35 10,clip,width=3.4cm,height=3.33cm]{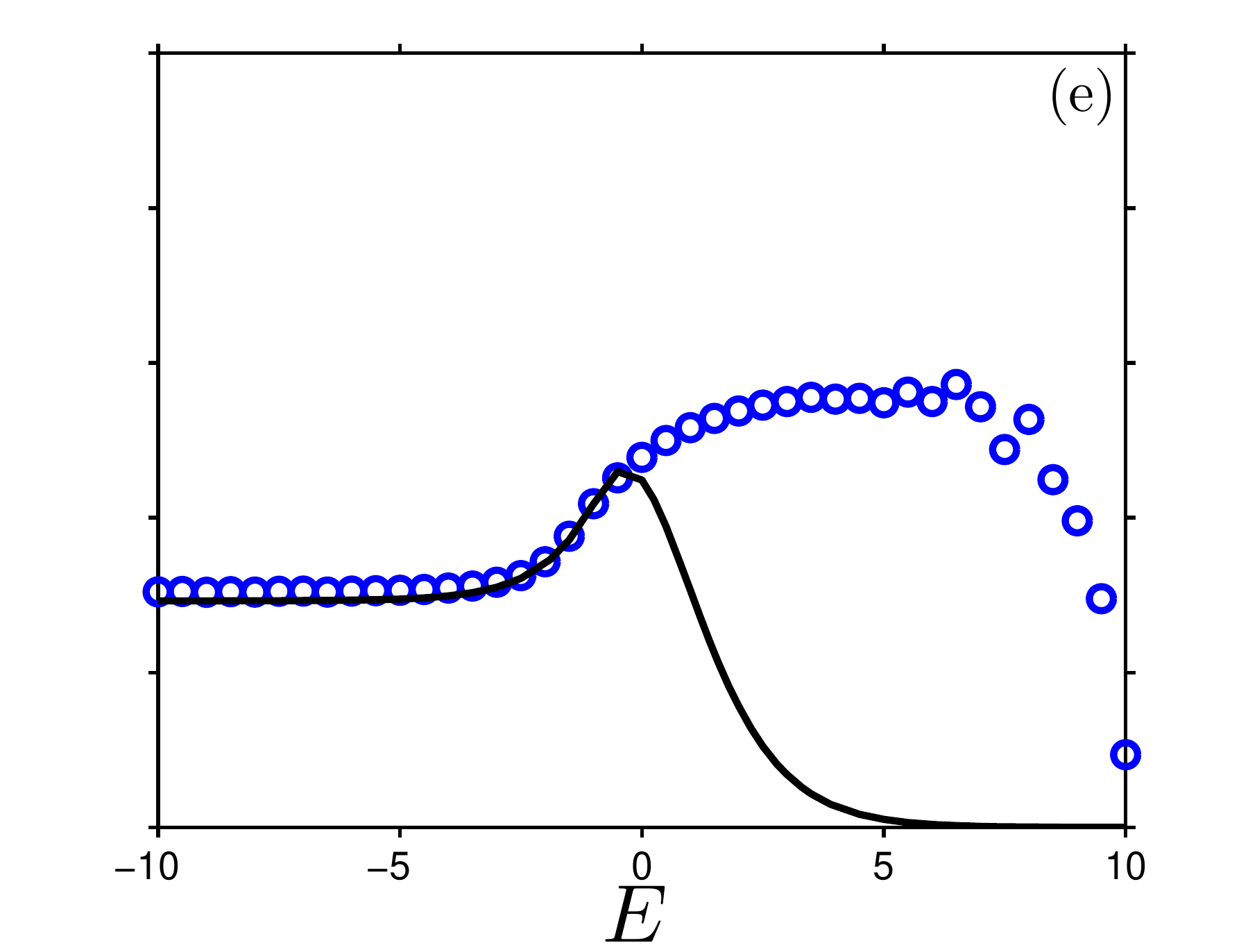}
\caption{Plots of maximal particle current ($J$) with respect to interaction energy ($E,\lambda^{-1})$ under the coupling strength $w = 0.2$ for different interaction splittings: (a) $\theta$ = 0; (b) $\theta$ = 0.25; (c) $\theta$ = 0.5; (d) $\theta$ = 0.75; (e) $\theta$ = 1. In simulations, $\alpha$ = 1, $\beta$ = 1 is utilised.}\label{fig6}
\end{center}
\end{figure*}
In a coupled interactive TASEP system, it is also important to discuss the effect of coupling rate on the phase diagram. Each curve in Fig. \ref{fig5} is traced by the position of the triple point corresponding to various coupling rate under a given interaction strength split symmetrically between $q$ and $r$. Note that the generalized theory correctly predicts the triple points for the special case of no interactions as simulation results coincides with the theoretical predictions (Fig. \ref{fig5}). For relatively weak attractive interactions, the triple points predicted from the theoretical approximation are in agreement with the simulation results, while for weak repulsive interactions, the location of the triple point agrees well only for relatively weak coupling $(w<0.25)$. It is found that the effect of coupling rate on multi-particle dynamics is symmetric with respect to interactions. In general, both simulation and theoretical results indicate that coupling between the lanes causes the shrinkage of HD region  irrespective of the presence of repulsive or attractive interaction. As when a particle feels obstruction in its own lane, it switches to other lane (with rate $w$) and thus diminishes the risk of traffic jam of particles and allows the smooth flow of particles.

To observe the effect for a range of interaction on maximal current of particles and
to validate whether the proposed theory overcomes the drawback of MFT in producing infinite bulk current for attractive interactions, particle flux in MC phase is computed as a function of interaction energy $E$ for different value of splitting parameter (Fig. ~\ref{fig6}). For repulsive interactions, a good agreement is found between theoretical and simulation results for any value of splitting parameter, except for $\theta = 0$ for which it holds qualitatively. While for attractive interaction, the results match only for smaller values of $\theta$. Also, particle flux behavior is not symmetric about $E=0$. Note that for $E>>1$, the maximal current vanishes for any value of $\theta$.

From the theoretical development and simulation results, one can see that as moving from very large repulsive to attractive interaction maximal particle current shows a non-monotonous behavior for all values of splitting parameter except for $\theta = 0$. It is rather a unimodal function, which firstly increases from a saturated positive value to its maximum with increments in $E$, then decreases rapidly, ultimately diminishing to zero as $E \rightarrow \infty$ for any value of splitting parameter $\theta \neq 0$ (Fig. \ref{fig6}).
Interestingly, it is found that in the coupled interactive two-channel system, the optimal value of the interaction energy corresponding to maximum flux does not belong to the case of no interactions. Rather it occurs for weak repulsive interactions for any value of splitting parameter (Fig. \ref{fig6}).

\begin{figure}
\begin{center}
\includegraphics[trim = 0 0 0 0,clip,width =7.6cm, height = 6.6cm]{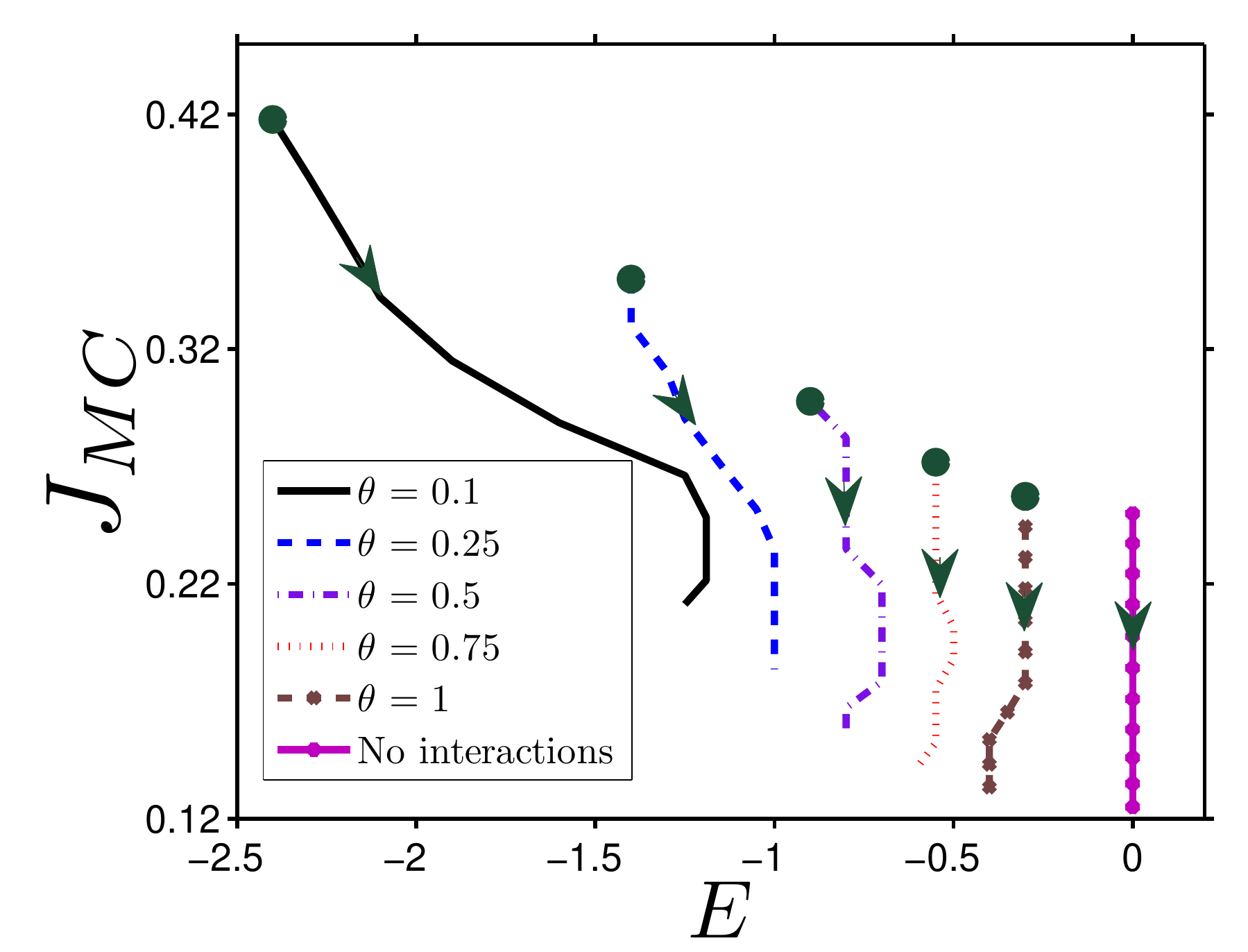}
\caption{Path traced by points representing optimal interaction strength ($E^*$)and its corresponding maximal particle current ($J_{MC}$) as a function of coupling rate $w$ for different $\theta$. The arrows on each curve point the direction of $w$ increasing from 0 to 1. Highlighted circle at each curve for the interactions indicates $E^*$ and corresponding maximal current for the case of single channel interacting TASEP model.}\label{fig7}
\end{center}
\end{figure}

To further investigate the effect of different coupling rates on the multi-particle interactive dynamics, the optimal interaction energy $(E^*)$ and its corresponding maximal particle current is plotted as a function of coupling rate $w$ in the phase plane $(E,J_{MC})$ for different $\theta$ (Fig. \ref{fig7}). It is found that whatever be the coupling rate, for a given parameter $\theta$ the optimal interaction strength corresponding to the maximal current of particles is a weak repulsive interaction strength. Also, irrespective of any value of splitting parameter $\theta$, the particle maximal current decreases with an increase in the coupling rate $w$. The optimal maximal current and the interaction strength obtained from simulation results agree quite well with the theoretical values shown in Fig.(\ref{fig7}) for all splitting values (except for $\theta\rightarrow 1)$. For any value of coupling rate $w$, the optimal current is found to be larger for smaller value of $\theta$. Most importantly, it is found that for weak splittings, $E^*$ decreases in magnitude with an increase in the coupling rate. Such behavior of $E^{*}$ is experimentally favorable.

%
 Experimentally, it is known that kinesins experience an interaction of range $(1.6\pm0.5) \lambda^{-1}$, which is crucial for maintaining the smooth flow of motors \cite{roos2008dynamic}.
The recent theoretical and numerical study on an interactive single channel TASEP model \cite{teimouri2015theoretical}, that takes only symmetric splitting of interaction,
argues the optimal interaction strength corresponding to the maximal current of particles to be $-3\lambda^{-1}$ which is in opposite regime to the experimentally known interaction strength. Later on, a refined theoretical study on the interactive single channel TASEP \cite{celis2015correlations}, called modified cluster mean field theory, takes the role of symmetry of interaction and claims the occurrence of $E^*$ in the range $-$(0.5-2)$\lambda^{-1}$ for most of the interaction splittings.

However, a single-channel and its theories cannot consider the combined effect of inter and intra-channel transitions, which affects the collective dynamics of motor proteins. As for $w = 0$, the developed generalized theory reduces to modified cluster mean field theory, the optimal interaction strength, $E^*$ for $w = 0$ is also highlighted by a circle in Fig. \ref{fig7} for the sake of comparison. The analysis of our results presented in figures \ref{fig6} and \ref{fig7} suggest that for weak interaction splittings, the possibility for the particle to change lane shifts the optimal interaction energy $(E^*)$ towards the experimentally found interactive strength of particles which is a result of great importance. Since in our model, we have incorporated only symmetric coupling between two interactive single-channel TASEP's, it is expected that if more realistic features of motor proteins such as their motion in a complex network, presence of three or more lanes, interaction with the open environment, backward stepping of motor proteins etc are considered, it might happen that kinesins working under the interaction energy $E = 1.6 \lambda^{-1}$ support maximal current as well.
The results in figures \ref{fig6} and \ref{fig7} also indicate that in spite of taking the effect of coupling in between the lanes, the small changes in interaction energy
may lead to large changes in particle collective behavior, which has been found experimentally important for maintaining robust cellular transport \cite{kolomeisky2013motor,teimouri2015theoretical,celis2015correlations}.
    \begin{figure*}
\begin{center}
\includegraphics[trim = 0 0 0 00,clip,width=4cm,height=3.4cm]{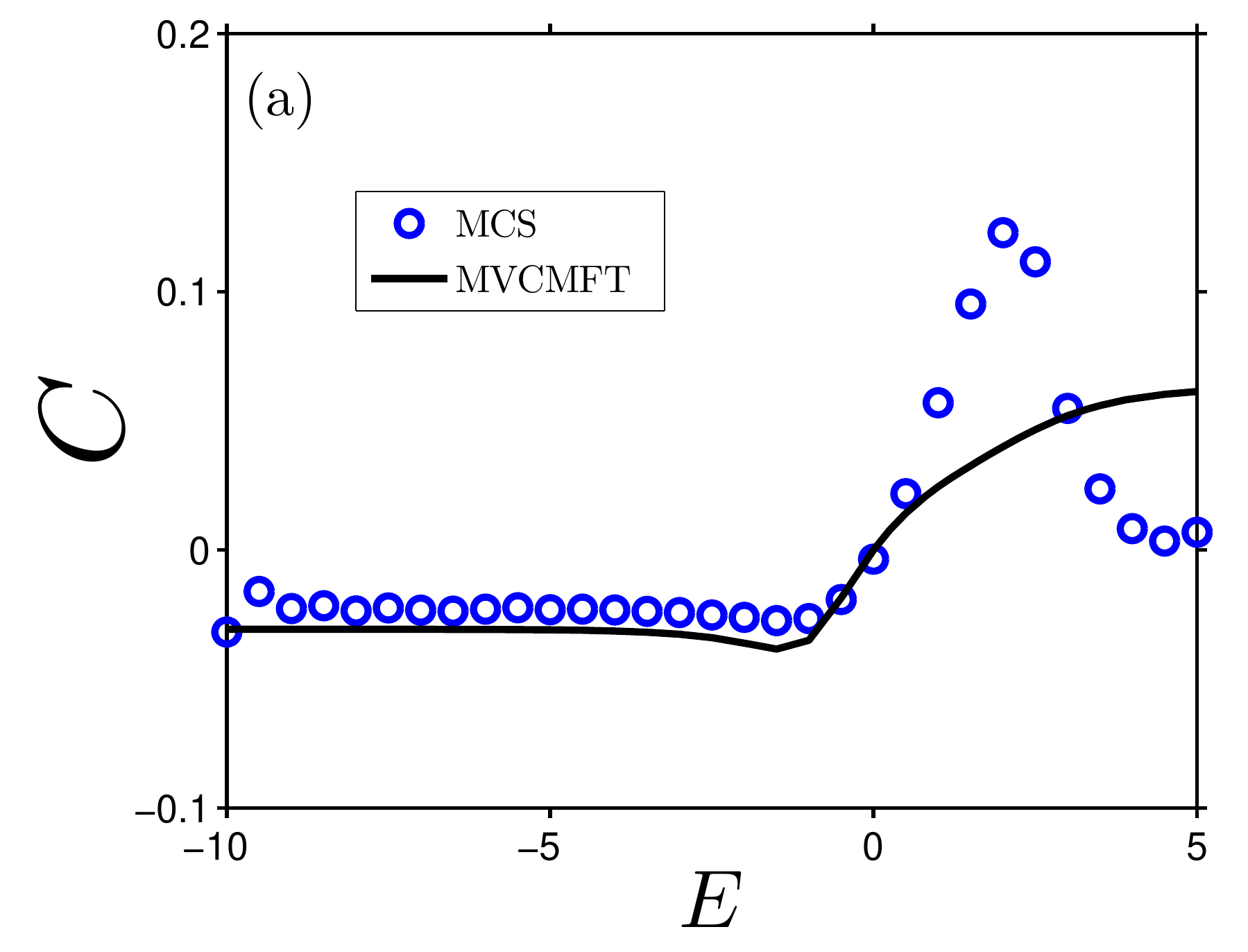}
\includegraphics[trim = 50 0 40 0,clip,width=3.24cm,height=3.43cm]{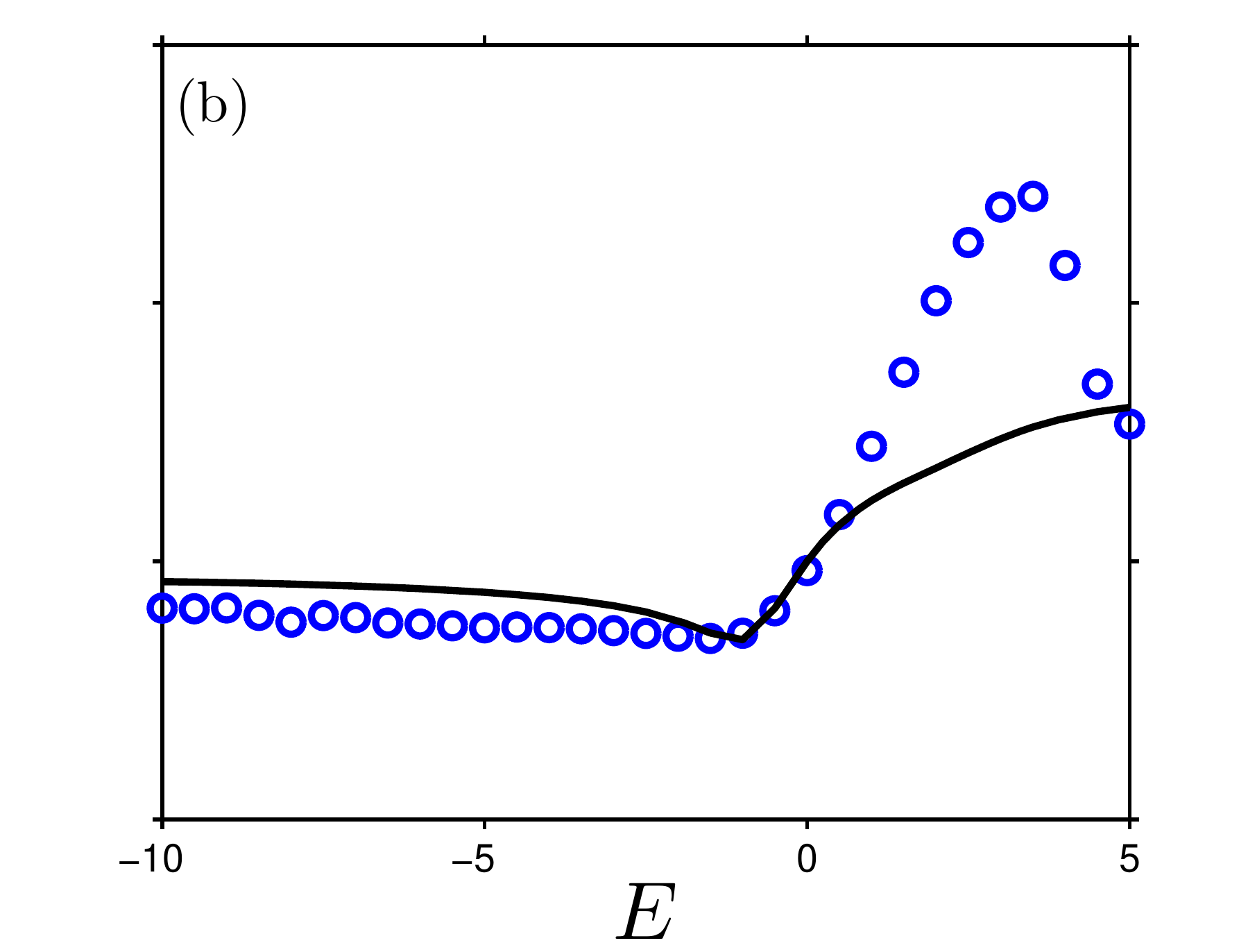}
\includegraphics[trim = 35 0 37 0,clip,width=3.42cm,height=3.41cm]{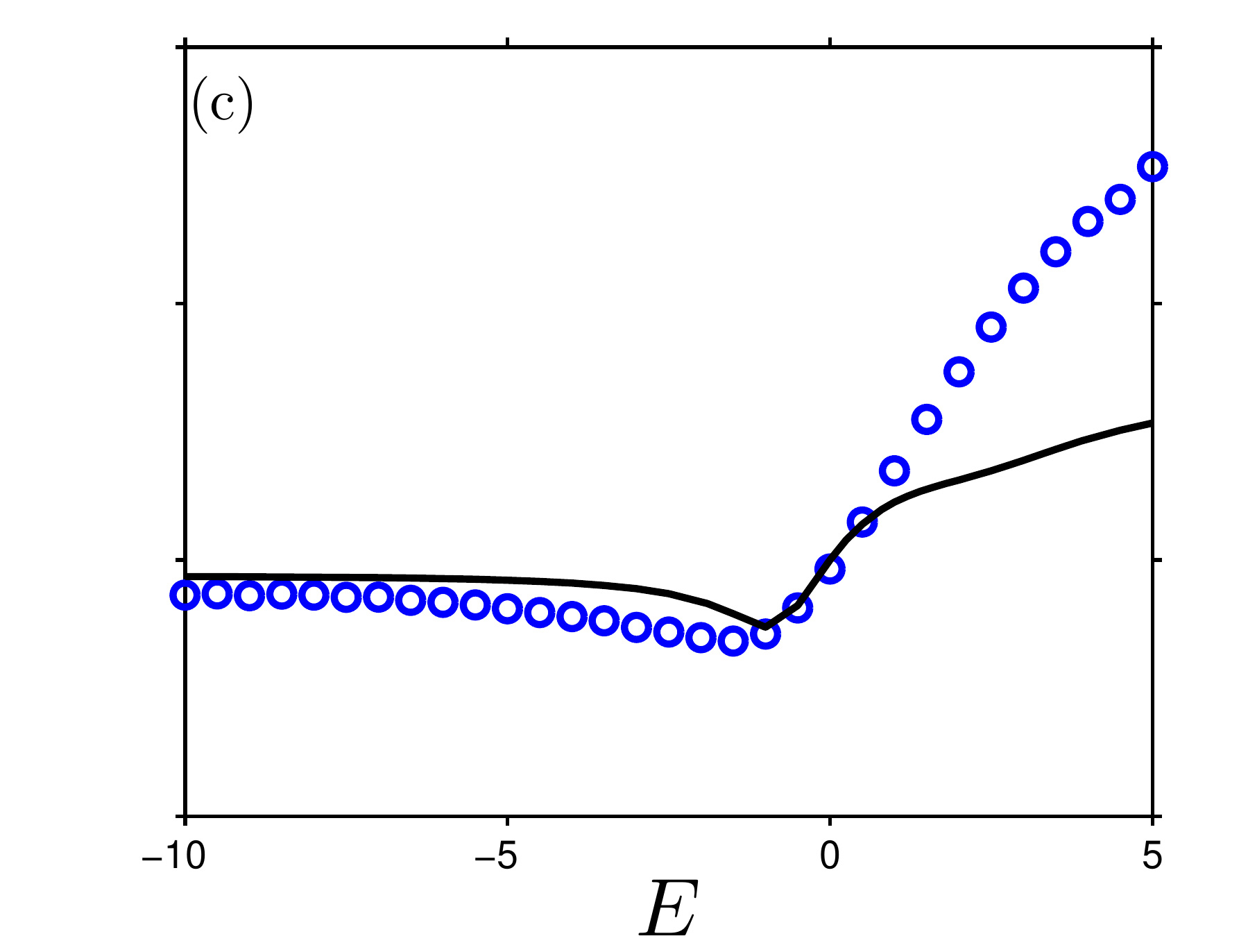}
\includegraphics[trim = 27 6 28 0,clip,width=3.5cm,height=3.36cm]{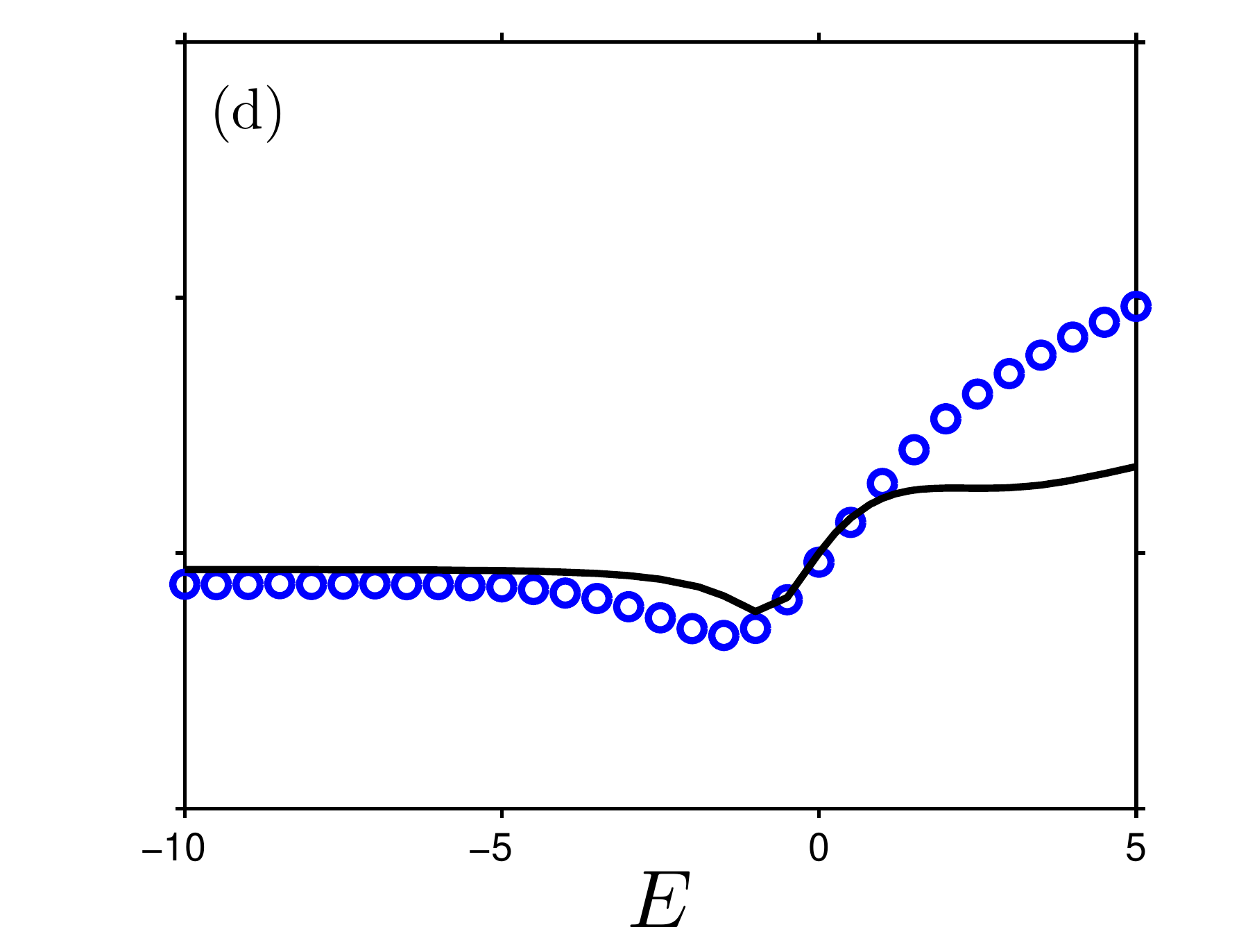}
\includegraphics[trim = 38 0 35 00,clip,width=3.38cm,height=3.32cm]{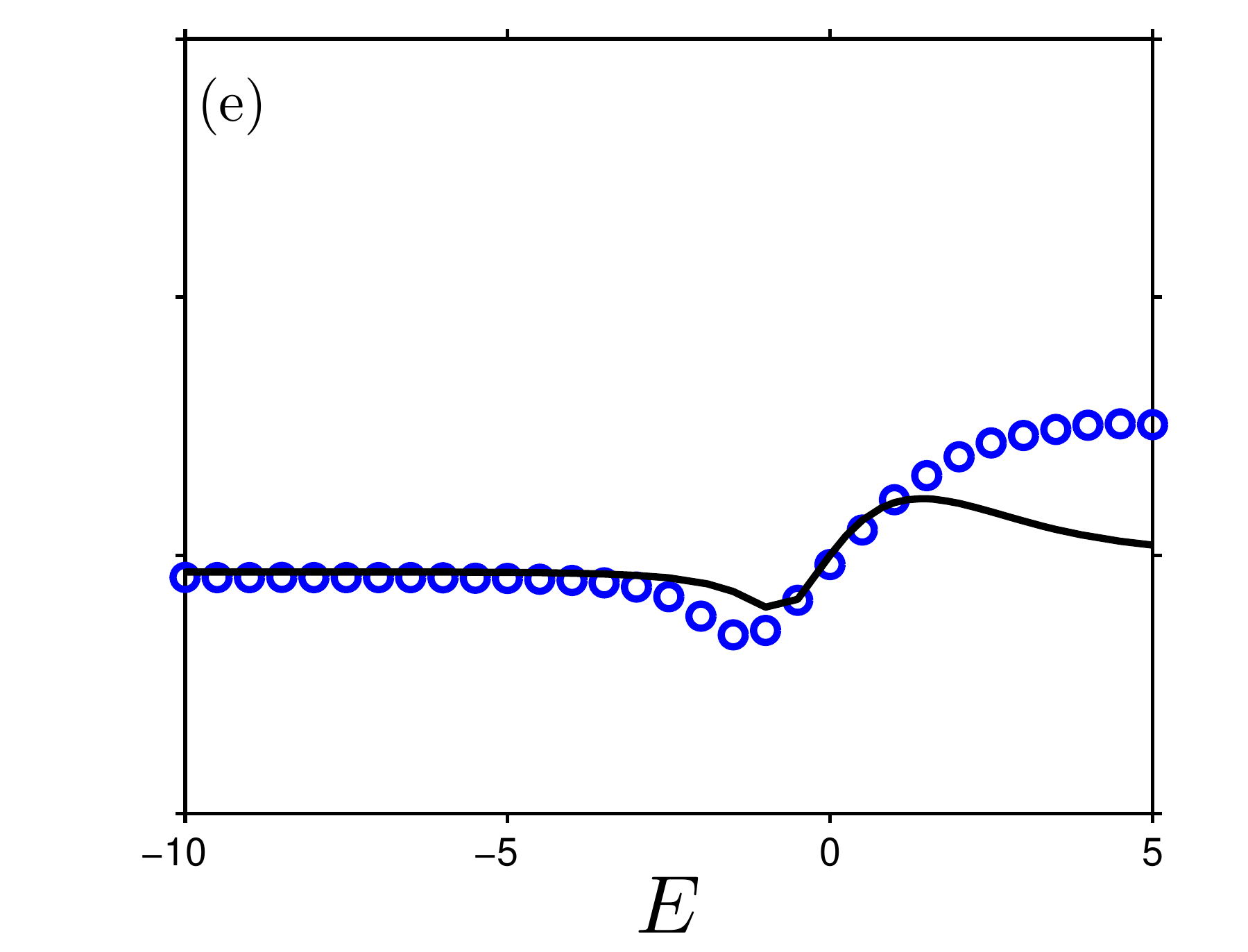}
\caption{Plots of correlation function ($C$) with respect to interaction energy ($E,\lambda^{-1})$ under the coupling strength $w = 0.2$ for: (a) $\theta$ = 0; (b) $\theta$ = 0.25; (c) $\theta$ = 0.5; (d) $\theta$ = 0.75; (e) $\theta$ = 1. In simulations, $\alpha$ = 1, $\beta$ = 1 is utilised.}\label{fig8}

\end{center}
\end{figure*}

As seen in our previous discussion that interactions of molecular motor in a coupled system significantly affect the particle dynamics, we now study the correlations to further investigate the dynamical properties of the system at the stationary state.
We define the two-point classical correlation function, where each point is considered as a vertical cluster, as
\begin{eqnarray}
\hspace{-0.5cm}C_{i} &= \langle\langle\tau_{i,l}\tau_{i,l'}\rangle\langle\tau_{i+1,l}\tau_{i+1,l'}\rangle\rangle - \langle\tau_{i,l}\tau_{i,l'}\rangle\langle\tau_{i+1,l}\tau_{i+1,l'}\rangle
\end{eqnarray}
where $i$ = 1, 2, $\cdots$, $N-1$,    and  $l \neq l' \in \lbrace1,2\rbrace$.\\

The one-point density function can be expressed as

\begin{eqnarray}
&\langle\tau_{i,l}\tau_{i,l'}\rangle = \sum_{\tau_{i,l}}\sum_{\tau_{i,l'}}\tau_{i,l}\tau_{i,l'}P(\tau_{i,l},\tau_{i,l'}),
\end{eqnarray}
which amounts to the density of fully filled vertical cluster, given by $V_3$.
The two-point density function is defined by
\begin{eqnarray}
\begin{aligned}
\langle\langle\tau_{i,l}\tau_{i,l'}\rangle\langle\tau_{i+1,l}\tau_{i+1,l'}\rangle\rangle & = \sum_{\tau_{i,l}}\sum_{\tau_{i,l'}}\sum_{\tau_{i+1,l}}\sum_{\tau_{i+1,l'}} \tau_{i,l}\tau_{i,l'}\tau_{i+1,l}\tau_{i+1,l'} \\
& \times P(\tau_{i,l},\tau_{i,l'},\tau_{i+1,l},\tau_{i+1,l'}).\label{eq32}\end{aligned}
\end{eqnarray}
Equation (\ref{eq32}) is the expected value of two neighboring vertical cluster and ultimately gives the probability for all four sites of the two-site vertical cluster to occupy simultaneously. Theoretically, this probability is approximated as $\frac{\eta V_3^2}{1-\rho + \eta\rho}$.

Thus, the correlation function for the coupled system can be expressed analytically in the closed form as
\begin{eqnarray}
C(\eta) = \frac{(\eta - 1)(1-\rho)V_3^2}{[1 + (\eta - 1)\rho]}.\label{eq31}
\end{eqnarray}

The above correlation function physically represents the extent by which the fully filled vertical cluster influences the occupancy of particles at its neighboring vertical cluster.
  When the probability of finding particles at $(i+1)^{th}$ vertical cluster is more, given that $i^{th}$ vertical cluster is fully filled, the correlation function takes positive value. This is corresponding to the case of attractive interaction, where particles  form large clusters. On the other hand, the function $C$ takes the negative values, for the case when presence of particles at $i^{th}$ vertical cluster reduce the chance to have particles at its neighboring vertical cluster. This occurs for the case of repulsive interaction when particles repel and there is less probability of finding two fully filled cluster together. Our theoretical expression for correlation function is in accordance with these physical considerations.

  For $\eta < 1$ i.e.  the case of repulsive interaction, $C$ yields negative values and approaches to $-V_3^2$ as $\eta \rightarrow 0$. 
For attractive interaction i.e. when $\eta > 1$, correlations are always positive and for a particular case of infinite attractions ($\eta \rightarrow \infty$), $C$ tends to extreme positive value $\frac{V_3^2(1-\rho)}{\rho}$. This positive and large value indicates the highest probability of getting  two fully filled vertical clusters together. When the occupancy of particles in two consecutive vertical clusters are independent for each other (in case of $\eta = 1)$, $C$ approaches towards zero.
It is worth to mention that effect of inter-channel interactions has appeared in $C$ through its natural inheritance in $V_3$. 


In Fig. \ref{fig8}, the correlation function obtained analytically as well as through simulations is plotted as a function of interaction energy $E$ for various value of splitting parameter $\theta$ and for a fixed coupling rate $w = 0.2$. It is clear from the figure that for repulsive interaction as well as weak attractive interaction, there is good consistency found between both theoretical and simulation findings whereas results match only qualitatively for relatively stronger attractive interactions. The reasons for above agreements are that for the case of repulsion particles generally form smaller cluster and influence locally, thus the correlations encountered are short-range and weak and can be captured by considering correlation between two neighboring vertical clusters. On the contrary for strong attractive interactions, particles at $i^{th}$ vertical cluster attract the particles to occupy its neighboring vertical cluster which further attract particles towards it leading to the formation of larger clusters. Thus, correlations developed are stronger and long-range which can not be justified by our approximation theory. Note that all correlation curves are plotted with respect to interaction energy $E$ instead of parameter $\eta$ for simplicity in comparing the results of one lane with the new discussed results. It is observed that comparative to the case of $w = 0$, coupling in between the lanes quantitatively decreases the correlations 
(\cite{celis2015correlations}, Fig.\ref{fig8}). Computed correlation curves in Fig. \ref{fig8} are for a fixed coupling strength $w = 0.2$ and it is checked that for other coupling strengths, the difference in correlation function curves is very minute and can be neglected within a relative error of $1\%.$

\section{Summary and Conclusions}
To summarize, we have developed a new generalized approach called modified vertical cluster mean field theory, to investigate the collective dynamics of interacting molecular motors that move along two parallel cytoskeletal filaments. The interactions, predicted experimentally, are incorporated in a two-channel symmetrically coupled original TASEP model by modifying its transition rates via fundamental thermodynamic arguments. It was observed that interactions bring correlations into the system, which can not be ignored. It causes the probability for any one-site vertical cluster to depend on the occupancy of its neighboring cluster. Our theory is based on approximating the probabilities of two neighboring one-site vertical clusters. The approach successfully predicts the correlations for repulsive and weak attractive interactions that bring short-range correlations, while it qualitatively predicts long-range correlation generated in case of stronger attractive interactions. The effect of inter-channel coupling on triple points is found to be symmetric for weak repulsive and attractive interactions.
The effect of different coupling strength and symmetry of interactions are observed on the optimal interaction strength ($E^*$) and its corresponding maximal particle current. When breaking of a bond is strongly affected by the interactions, (i.e. $\theta$ is smaller), an increment in the coupling rate shifts the optimal interaction strength for maximal current toward the experimentally predicted energy for a rich flow of motor proteins. However, when the breaking of a bond is weakly influenced (i.e. $\theta$ is larger), an increment in the coupling rate has a minor affect on the optimal interactive strength $E^*$. 
The coupling in between both the channels decreases the magnitude of correlations into the system as compared to the case of one-channel interacting TASEP system. All results predicted from the proposed theory are matched with extensive Monte-Carlo simulations.

From the implication of our results for kinesin motor proteins, we conclude that implementing more realistic features of motor proteins such as interactions with more than one-channel and open-environment at all sites of a channel, limited number of motor proteins at the surroundings, back movement of motor proteins, etc gives more justification to the experimental expectations on kinesins. Our theory can easily be extended to take these features into account for a better understanding of motor protein behavior.
\appendix
\section{ Expression for bulk current using VCMF approach:}
 The Vertical cluster mean field approach is utilized to compute the bulk current for all eight possible configuration (shown in Fig. \ref{fig2}) that contribute to the particle movement in bulk. The total bulk current in the system is an algebraic sum of currents from all of these configurations. We first compute the particle current corresponding to configurations in Fig. \ref{fig2}(a-d) that doesn't involve the coupling parameter $w$ i.e. the case when the particle can not hop vertically. We denote total bulk currents from these four configurations as
\begin{eqnarray}
 J_1 = J_a + J_b + J_c + J_d, \label{A.1}
\end{eqnarray}
where $J_a$, $J_{b}$, $J_c$, and $J_{d}$ are particle currents corresponding to configuration (a), (b), (c), and (d), respectively, given as
\begin{eqnarray}
\begin{aligned}
J_{a} &= V_3(V_0 + V_2)^3, \label{A.2}\\
J_{b} &= qV_3(V_1+V_3)(V_0+V_2)^2, \label{A.3}\\
J_c &= rV_3(V_1+V_3)(V_0+V_2)^2, \label{A.4}\\
\hspace{-2.5cm}\mbox{and}~~~~~~~~~~~~~ J_d &= V_3(V_0+V_2)(V_1+V_3)^2,\label{A.5}
\end{aligned}
\end{eqnarray}
%
%
thus implying,
\begin{eqnarray}
J_1 = V_3(1 - \rho)[(1-\rho)^2 + (q+r)\rho(1-\rho) + \rho^2].\label{A.6}
\end{eqnarray}
Here $\rho$ = $V_1 + V_3$ denotes the particle bulk density.
Similarly particle bulk current for the configurations in Fig. \ref{fig2}(e-h) which involves the role of parameter $w$ i.e. when lane switching is possible, can be computed as
\begin{eqnarray}
\begin{aligned}
J_{e} &= (1-w)V_1(V_0 + V_2)^3, \label{A.7}\\
J_{f} &= q(1-w)V_1(V_1+V_3)(V_0+V_2)^2, \label{A.8}\\
J_g &= r(1-w)V_1(V_1+V_3)(V_0+V_2)^2,  \label{A.9} \\
\hspace{-3.8cm}\mbox{and}~~~~~~~~~J_h &= (1-w)V_1(V_0+V_2)(V_1+V_3)^2,\label{A.10}\end{aligned}
\end{eqnarray}
where  $J_e$, $J_{f}$, $J_g$, and $J_{h}$ denote particle currents corresponding to configuration (e), (f), (g), and (h), respectively. The total bulk currents for the above four configurations is expressed as
\begin{eqnarray}
\begin{aligned}
J_2 &= J_e + J_f +J_g +J_h \\
    &= (1-w)V_1(1 - \rho)[(1-\rho)^2 + (q+r)\rho(1-\rho) + \rho^2].
    \label{A.7}
\end{aligned}    
\end{eqnarray}
The overall bulk current per channel is
\begin{eqnarray}
\begin{aligned}
J_{bulk}^{VCMF} &= J_1 + J_2\\
&=(V_3 + (1-w)V_1)(1-\rho)[(1-\rho)^2 \\
&\hspace{0.95cm} +(q+r)\rho(1-\rho) + \rho^2].\label{A.8}\end{aligned}
\end{eqnarray}
The rate of formation of particle cluster, $q = \eta^{\theta}$, approaches infinity for very large attractive interactions ($\eta \gg 1)$ as $\theta > 0$. This causes bulk current to increase without any limit which in contrast to the physical implication that under large attractive strength particle current should ultimately die out. Similarly, the rate of deformation of cluster, $r = \eta^{(\theta - 1)} $, tends to infinity for $\eta \rightarrow 0 $ and $\theta < 1$.
\section{ Monte-Carlo Simulations}
Due to approximate nature of our method in calculating the effect of interactions and correlations, we validate the results obtained from the given approximate theoretical method with extensive Monte Carlo (MC) Simulations. Random-Sequential update rules are adopted. For a single Monte-Carlo step, first a lattice is randomly chosen with equal probability. To avoid any finite-size and boundary effects both lattices are considered to be of size $N$ = 1000 unless otherwise mentioned. The results have been verified by taking large lattice size of $L =5000$. The simulation starts from a random initial distribution of particles on both the lattices and system evolved for $10^{9}$ to $10^{10}$ time steps to ensure steady state condition.
To compute density and particle current at steady state, an average of the last $80\%$ of the steps has been taken. In constructing phase diagrams, density profiles are compared with a precision of 0.01 and for calculating phase boundaries, error estimated in comparing currents is less than $1\%$. Our predicted theoretical results fit well with the simulation results.

\section*{References}

\begin{thebibliography}{10}





\bibitem{alberts2002molecular}
B~Alberts, A~Johnson, J~Lewis, P~Walter, M~Raff, and K~Roberts.
\newblock Molecular biology of the cell 4th edition: International student
  edition, 2002.

\bibitem{bray2001cell}
Dennis Bray.
\newblock {\em Cell movements: from molecules to motility}.
\newblock Garland Science, 2001.



\bibitem{howard2001mechanics}
Jonathon Howard et~al.
\newblock Mechanics of motor proteins and the cytoskeleton.
\newblock 2001.


\bibitem{kolomeisky2007molecular}
Kolomeisky, Anatoly B and Fisher, Michael E.
\newblock Molecular motors: a theorist's perspective.
\newblock {\em Annu. Rev. Phys. Chem.}, 58:675--695, 2007.

\bibitem{chowdhury2013stochastic}
Debashish Chowdhury.
\newblock Stochastic mechano-chemical kinetics of molecular motors: a
  multidisciplinary enterprise from a physicist?s perspective.
\newblock {\em Physics Reports}, 529(1):1--197, 2013.


\bibitem{kolomeisky2013motor}
Anatoly~B Kolomeisky.
\newblock Motor proteins and molecular motors: how to operate machines at the
  nanoscale.
\newblock {\em Journal of Physics: Condensed Matter}, 25(46):463101, 2013.

\bibitem{kolomeisky2015motor}
Anatoly~B Kolomeisky.
\newblock {\em Motor Proteins and Molecular Motors}.
\newblock CRC Press, 2015.



\bibitem{schliwa2003molecular}
Manfred Schliwa and G{\"u}nther Woehlke.
\newblock Molecular motors.
\newblock {\em Nature}, 422(6933):759--765, 2003.


\bibitem{veigel2011moving}
Claudia Veigel and Christoph~F Schmidt.
\newblock Moving into the cell: single-molecule studies of molecular motors in
  complex environments.
\newblock {\em Nature Reviews Molecular Cell Biology}, 12(3):163--176, 2011.


\bibitem{ally2009opposite}
Shabeen Ally, Adam~G Larson, Kari Barlan, Sarah~E Rice, and Vladimir~I Gelfand.
\newblock Opposite-polarity motors activate one another to trigger cargo
  transport in live cells.
\newblock {\em The Journal of cell biology}, 187(7):1071--1082, 2009.


\bibitem{driver2011productive}
Jonathan~W Driver, D~Kenneth Jamison, Karthik Uppulury, Arthur~R Rogers,
  Anatoly~B Kolomeisky, and Michael~R Diehl.
\newblock Productive cooperation among processive motors depends inversely on
  their mechanochemical efficiency.
\newblock {\em Biophysical journal}, 101(2):386--395, 2011.

\bibitem{driver2010coupling}
Jonathan~W Driver, Arthur~R Rogers, D~Kenneth Jamison, Rahul~K Das, Anatoly~B
  Kolomeisky, and Michael~R Diehl.
\newblock Coupling between motor proteins determines dynamic behaviors of motor
  protein assemblies.
\newblock {\em Physical Chemistry Chemical Physics}, 12(35):10398--10405, 2010.

\bibitem{neri2013exclusion}
Izaak Neri, Norbert Kern, and Andrea Parmeggiani.
\newblock Exclusion processes on networks as models for cytoskeletal transport.
\newblock {\em New Journal of Physics}, 15(8):085005, 2013.


\bibitem{roos2008dynamic}
Wouter~H Roos, Fabien Montel, Joachim~P Spatz, Patricia Bassereau, Giovanni
  Cappello, et~al.
\newblock Dynamic kinesin-1 clustering on microtubules due to mutually
  attractive interactions.
\newblock {\em Physical biology}, 5(4):046004, 2008.


\bibitem{seitz2006processive}
Arne Seitz and Thomas Surrey.
\newblock Processive movement of single kinesins on crowded microtubules
  visualized using quantum dots.
\newblock {\em The EMBO journal}, 25(2):267--277, 2006.


\bibitem{uppulury2012interplay}
Karthik Uppulury, Artem~K Efremov, Jonathan~W Driver, D~Kenneth Jamison,
  Michael~R Diehl, and Anatoly~B Kolomeisky.
\newblock How the interplay between mechanical and nonmechanical interactions
  affects multiple kinesin dynamics.
\newblock {\em The Journal of Physical Chemistry B}, 116(30):8846--8855, 2012.


\bibitem{macdonald1968kinetics}
Carolyn~T MacDonald, Julian~H Gibbs, and Allen~C Pipkin.
\newblock Kinetics of biopolymerization on nucleic acid templates.
\newblock {\em Biopolymers}, 6(1):1--25, 1968.







%
%
%

\bibitem{belitsky2001cellular}
Belitsky, V and Krug, J and Neves, E Jordao and Sch{\"u}tz, GM.
\newblock A cellular automaton model for two-lane traffic.
\newblock {\em Journal of Statistical Physics}, 107(5-6):945--971, 2001.


\bibitem{chowdhury2008traffic}
Chowdhury, Debashish and Garai, Ashok and Wang, Jian-Sheng.
\newblock Traffic of single-headed motor proteins KIF1A: effects of lane changing.
\newblock {\em Physical Review E}, 77(5):050902, 2008.

\bibitem{widom1991repton}
B~Widom, JL~Viovy, and AD~Defontaines.
\newblock Repton model of gel electrophoresis and diffusion.
\newblock {\em Journal de Physique I}, 1(12):1759--1784, 1991.







\bibitem{campas2006collective}
O~Campas, Y~Kafri, KB~Zeldovich, J~Casademunt, and J-F Joanny.
\newblock Collective dynamics of interacting molecular motors.
\newblock {\em Physical review letters}, 97(3):038101, 2006.


\bibitem{klumpp2004phase}
Klumpp, Stefan and Lipowsky, Reinhard.
\newblock Phase transitions in systems with two species of molecular motors.
\newblock {\em EPL (Europhysics Letters)}, 66(1):90, 2004.
\bibitem{pinkoviezky2013modelling}
Itai Pinkoviezky and Nir~S Gov.
\newblock Modelling interacting molecular motors with an internal degree of
  freedom.
\newblock {\em New Journal of Physics}, 15(2):025009, 2013.


\bibitem{slanina2008interaction}
Franti{\v{s}}ek Slanina.
\newblock Interaction of molecular motors can enhance their efficiency.
\newblock {\em EPL (Europhysics Letters)}, 84(5):50009, 2008.



\bibitem{teimouri2015theoretical}
Hamid Teimouri, Anatoly~B Kolomeisky, and Kareem Mehrabiani.
\newblock Theoretical analysis of dynamic processes for interacting molecular
  motors.
\newblock {\em Journal of Physics A: Mathematical and Theoretical},
  48(6):065001, 2015.


\bibitem{celis2015correlations}
Daniel Celis-Garza, Hamid Teimouri, and Anatoly~B Kolomeisky.
\newblock Correlations and symmetry of interactions influence collective
  dynamics of molecular motors.
\newblock {\em Journal of Statistical Mechanics: Theory and Experiment},
  2015(4):P04013, 2015.



\bibitem{hao2016exponential}
Hao, Qing-Yi and Jiang, Rui and Hu, Mao-Bin and Jia, Bin and Wang, Wen-Xu.
\newblock Exponential decay of spatial correlation in driven diffusive system: A universal feature of macroscopic homogeneous state.
\newblock {\em Scientific reports},
 6(19652):19652, 2016.




\bibitem{hao2016theoretical}
Hao, Qing-Yi and Chen, Zhe and Sun, Xiao-Yan and Liu, Bing-Bing and Wu, Chao-Yun.
\newblock Theoretical analysis and simulation for a facilitated asymmetric exclusion process.
\newblock {\em Physical Review E}, 94(2):022113,
  2016.

\bibitem{chou2011non}
Chou, T and Mallick, K and Zia, RKP.
\newblock Non-equilibrium statistical mechanics: from a paradigmatic model to biological transport.
\newblock {\em Reports on progress in physics}, 74(11):116601,
  2011.


\bibitem{dong2012entrainment}
Dong, Jiajia and Klumpp, Stefan and Zia, Royce KP.
\newblock Entrainment and unit velocity: Surprises in an accelerated exclusion process.
\newblock {\em Physical review letters}, 109(13):130602,
  2012.


\bibitem{shaebani2014anomalous}
M~Reza Shaebani, Zeinab Sadjadi, Igor~M Sokolov, Heiko Rieger, and Ludger
  Santen.
\newblock Anomalous diffusion of self-propelled particles in directed random
  environments.
\newblock {\em Physical Review E}, 90(3):030701, 2014.



\bibitem{pronina2004two}
Ekaterina Pronina and Anatoly~B Kolomeisky.
\newblock Two-channel totally asymmetric simple exclusion processes.
\newblock {\em Journal of Physics A: Mathematical and General}, 37(42):9907,
  2004.

\bibitem{pronina2006asymmetric}
Ekaterina Pronina and Anatoly~B Kolomeisky.
\newblock Asymmetric coupling in two-channel simple exclusion processes.
\newblock {\em Physica A: Statistical Mechanics and its Applications},
  372(1):12--21, 2006.

\bibitem{juhasz2007weakly}
Robert Juhasz.
\newblock Weakly coupled, antiparallel, totally asymmetric simple exclusion
  processes.
\newblock {\em Physical Review E}, 76(2):021117, 2007.


\bibitem{popkov2004hydrodynamic}
V~Popkov and M~Salerno.
\newblock Hydrodynamic limit of multichain driven diffusive models.
\newblock {\em Physical Review E}, 69(4):046103, 2004.

\bibitem{gupta2013coupling}
Arvind~Kumar Gupta and Isha Dhiman.
\newblock Coupling of two asymmetric exclusion processes with open boundaries.
\newblock {\em Physica A: Statistical Mechanics and its Applications},
  392(24):6314--6329, 2013.


\bibitem{derrida1992exact}
Bernard Derrida, Eytan Domany, and David Mukamel.
\newblock An exact solution of a one-dimensional asymmetric exclusion model
  with open boundaries.
\newblock {\em Journal of Statistical Physics}, 69(3-4):667--687, 1992.

\bibitem{derrida1993exact}
Bernard Derrida, Martin~R Evans, Vincent Hakim, and Vincent Pasquier.
\newblock Exact solution of a 1d asymmetric exclusion model using a matrix
  formulation.
\newblock {\em Journal of Physics A: Mathematical and General}, 26(7):1493,
  1993.






\bibitem{derrida1999bethe}
B~Derrida and MR~Evans.
\newblock Bethe ansatz solution for a defect particle in the asymmetric
  exclusion process.
\newblock {\em Journal of Physics A: Mathematical and General}, 32(26):4833,
  1999.






\bibitem{lakatos2003totally}
Greg Lakatos and Tom Chou.
\newblock Totally asymmetric exclusion processes with particles of arbitrary
  size.
\newblock {\em Journal of Physics A: Mathematical and General}, 36(8):2027,
  2003.







\end{thebibliography}

  \end{document}